# ETTORE MAJORANA :
## Introduzione a vita & opere

**Erasmo RECAMI**

*Facoltà di Ingegneria, Università statale di Bergamo, Bergamo, Italy,*
*and INFN-Sezione di Milano, Milan, Italy.*

2009

***An Abstract in English:***

*This agile small book (in Italian) is an introduction to the life and work of Ettore Majorana, the brightest Italian theoretical physicist of the XX century, regarded by Enrico Fermi as the best theoretician of his time in the world.*

*It consists of ten smooth chapters, only few of them being moderately technical, and of three Appendices: (i) on his inaugural lecture of physics at the University of Naples; (ii) on some notes of his about the role of the statistical laws in physics and in the social sciences; and (iii) with a recapitulatory biographic outline.*

*The present material is mainly taken from our book* Il Caso Majorana: Epistolario, Documenti, Testimonianze *, first published (1987, 1991) by Mondadori, Milan, and presently (2000, 2002, 2008) published by Di Renzo Editore, Rome. That book presents practically all the serious documents Existing on Majorana's life & work [indeed, almost all the biographical documents have been discovered or collected, during a few decades, by the present author, who was the first to publish them]. We address to such a book (c/o* www.direnzo.it, *"Arcobaleno" series) all the readers interested in more and deeper information.*

## Prefazione

Chi scrive ha avuto l'occasione, dal 1970 circa e fino a pochissimi anni fa, di scoprire o raccogliere –e per primo pubblicare—tutti i documenti seri noti su vita e opere del catanese Ettore Majorana, il più brillante fisico teorico italiano dello scorso secolo, paragonato da Enrico Fermi a geni "come Galileo e Newton". [Pochi anni fa il "testimone" (in anni già lontani nelle mani di Edoardo Amaldi) è passato in parte dalle mie alle mani del collaboratore Salvatore Esposito, di Napoli].

Nei primissimi anni settanta lo scrivente trasmise copia di una parte di detti documenti a Leonardo Sciascia, che ne fece uso per la scrittura del suo famoso saggio sulla Scomparsa di Majorana (come noto, il Majorana scomparve misteriosamente da Napoli, nel 1938, all'età di trentun anni).

La *summa* biografica è costituita da un nostro volume, scritto anche a nome di Maria Majorana, indimenticabile sorella di Ettore, e della di lui ex-cognata Nunni Cirino ved. Luciano Majorana: entrambe meravigliose amiche purtroppo scomparse. Detto volume apparve nel 1987 e 1991 nella serie Oscar Mondadori, Milano, e nei più recenti anni 2000-2002-2008 per i tipi della Di Renzo Editore, Roma; e ad esso si rimanda per chi volesse approfondire il tema (`www.direnzo.it`, collana "Arcobaleno").

Abbiamo pensato però a un volumetto più agile, con una decina di scorrevoli capitoli, solo *pochi* dei quali *moderatamente* tecnici; oltre ad alcune Appendici, pure di interesse soprattutto culturale.

Il volumetto a cui si pensava eccolo: è questo. Si basa in parte sulla numerosa serie di articoli che durante il 2006 (anno del centenario della nascita del Nostro) il quotidiano di Catania, "La Sicilia", volle da me; sono grato al suo responsabile della Cultura, Salvatore Scalia: senza l'obbligo da lui "impostomi" di redigere tanti articoli giornalistici, il presente libro non esisterebbe. Sono pure grato alle varie Riviste, e ai loro responsabili, che mi hanno stimolato a scrivere altre pagine divulgative; e soprattutto all'editore Sante Di Renzo per avermi permesso di riprendere vari brani dal corposo libro da me presso di lui pubblicato.

Ovviamente i miei collaboratori ed io abbiamo pubblicato numerosi articoli scientifici, e libri (presso editori internazionali come Springer e Kluwer, e nazionali quali Zanichelli e Bibliopolis), sull'opera del Majorana: soprattutto sui manoscritti scientifici da lui lasciati inediti; ma a tutto ciò accenneremo solo di sfuggita. Chi volesse sapere di più intorno a vita e opere del Majorana potrà consultare, oltre al libro sopra menzionato, la nostra home-page `www.unibg.it/recami` (o anche il settore *physics* dell'archivio usato dai fisici per mettere in rete i loro manoscritti prima della pubblicazione, `www.arxiv.org`, digitando ad esempio "recami" sulla riga Autori).

Erasmo Recami (*)

(*) Fisico teorico, biografo di Majorana, e past-president della associazione "Amici di Sciascia": `recami@mi.infn.it` .

CAPITOLO PRIMO

**ETTORE MAJORANA: PRESENTAZIONE**

> Chissà quanti sono come me, nelle
> mie stesse condizioni, fratelli miei.
> Si lascia il cappello e la giacca,
> con una lettera in tasca, sul parapetto
> d'uno ponte, su un fiume; e poi,
> invece di buttarsi giù, si va via
> tranquillamente: in America o altrove.
> *L. Pirandello ("Il fu Mattia Pascal", 1904).*

**La fama**
La fama di Ettore Majorana, ovvia per gli specialisti, può solidamente appoggiarsi a testimonianze come la seguente, dovuta alla memore penna di Giuseppe Cocconi. Invitato da Edoardo Amaldi (il primo storico di Ettore Majorana) dal CERN gli scrive:

<<Ginevra, 1965 Luglio 18 --- Caro Amaldi, [...] mentre mi trovavo con Fermi arrivò la notizia della scomparsa da Napoli del Majorana. Mi ricordo che Fermi si dette da fare telefonando da varie parti sinché, dopo alcuni giorni, si ebbe l'impressione che non lo si sarebbe ritrovato più. Fu allora che Fermi, cercando di farmi capire che cosa significasse tale perdita, si espresse in modo alquanto insolito, lui che era così serenamente severo quando si trattava di giudicare il prossimo. Ed a questo punto vorrei ripetere le sue parole, così come da allora me le sento risuonare nella memoria: *"Perché, vede, al mondo ci sono varie categorie di scienziati; gente di secondo e terzo rango, che fan del loro meglio ma non vanno molto lontano. C'è anche gente di primo rango, che arriva a scoperte di grande importanza, fondamentali per lo sviluppo della scienza* (e qui ho netta l'impressione che in quella categoria volesse mettere se stesso). *Ma poi ci sono i geni, come Galileo e Newton. Ebbene, Ettore era uno di quelli. Majorana aveva quel che nessun altro al mondo ha..."*>>.

Enrico Fermi è uno dei maggiori fisici della nostra epoca: per quello che ha fatto nel 1942 a Chicago (con la costruzione della prima "pila atomica") il suo nome diverrà forse leggendario quanto quello di Prometeo; ed è stato l'ultimo grande esempio di fisico sia teorico sia sperimentale: *eppure* riconosceva la superiorità del Majorana nella fisica teorica. Egli si espresse in maniera per lui insolita anche in un'altra occasione: il 27 luglio 1938 (dopo la scomparsa di Majorana, avvenuta il sabato 26 marzo 1938), scrivendo quanto segue da Roma al primo ministro Mussolini onde chiedere una intensificazione delle ricerche di Ettore:

<<Io non esito a dichiararVi, e non lo dico quale espressione iperbolica, che fra tutti gli studiosi italiani e stranieri che ho avuto occasione di avvicinare il Majorana è fra tutti quello che per profondità di ingegno mi ha maggiormente colpito>>.

Se si pensa che Fermi conosceva praticamente tutti gli scienziati più eminenti del suo tempo, questa sua asserzione deve fare meditare: forse Ettore Majorana era il più grande teorico non solo d'Italia, ma del mondo: almeno del suo tempo.

E, infatti, un testimone diretto, Bruno Pontecorvo, affermò: <<Qualche tempo dopo l'ingresso nel gruppo di Fermi, Majorana possedeva già una erudizione tale ed aveva raggiunto un tale livello di comprensione della fisica da potere parlare con Fermi di problemi scientifici da pari a pari. Lo stesso Fermi lo riteneva il più grande fisico teorico dei nostri tempi. Spesso ne rimaneva stupito [...]. Ricordo esattamente queste parole di Fermi: "Se un problema è già posto, nessuno al mondo lo può risolvere meglio di Majorana".>>

Il mito della scomparsa ha contribuito a dare a Majorana, quindi, null'altro  che la notorietà  che gli spettava, per essere egli davvero un genio: e di una genialità precorritrice dei tempi. Anzi, così come avviene quando è vera, la sua fama è cresciuta e cresce col tempo, anche tra i suoi colleghi fisici.

Dicevamo che Fermi è stato forse l'ultimo --e straordinario-- esempio di grande teorico e contemporaneamente di grande sperimentale. Majorana era invece un teorico puro, anzi (per dirla con le stesse parole di Fermi, nel prosieguo del suo scritto a Mussolini) Ettore aveva al massimo grado quel raro complesso di attitudini che formano il fisico teorico di gran classe. Ettore "portava" la scienza, come ha detto Leonardo Sciascia: portava, anzi, la fisica teorica. Da un lato, quindi, non aveva alcuna propensione per le attività sperimentali (neanche costretto, per intenderci, avrebbe mai potuto recare contributi concreti a progetti come quello della costruzione tecnologica della bomba atomica). Dall'altro lato, però, sapeva calarsi a profondità insuperate nella sostanza dei fenomeni fisici, leggendovi eleganti simmetrie e nuove potenti strutture matematiche, o scoprendovi raffinate leggi. La sua acutezza lo portava a vedere al di là dei colleghi: ad essere cioè un pioniere. Perfino i suoi appunti di studio --redatti ciascuno in circa un anno, a partire dagli inizi del 1927-- sono un modello di scelta dell'essenziale, sinteticità e soprattutto originalità. Essi, noti come i *Volumetti*, sono stati da noi pubblicati nel 2003, in traduzione inglese e in forma digitata elettronicamente (512 pagine!) per i tipi della Kluwer Academic Press, di Dordrecht e Boston; e nella lingua originale italiana presso la Zanichelli, Bologna (2006).

Anticipiamo che Majorana, passato a fisica agli inizi del '28, si laureò con Fermi il 6 luglio 1929; e conseguì la libera docenza in fisica teorica il 12 novembre 1932.

**Da Galileo a Fermi**

Per dare un'idea di cosa abbia significato per la cultura e la scienza italiane l'attività romana di Fermi e del suo gruppo (con questo senza dimenticare la contemporanea attività di altri gruppi, in primis quello di Firenze), ricordiamo che la fisica italiana già una volta aveva conquistato una posizione di eccellenza a livello internazionale: con Galileo. Ma la condanna da parte della Chiesa (22 luglio 1633) --che, considerati i tempi, non ebbe in fondo conseguenze molto gravi per Galileo-- risultò disastrosa per la "scuola" galileana, la quale avrebbe potuto continuare ad essere la prima del mondo. Il vasto e promettente movimento scientifico creato dal Galilei venne colpito alla radice dalla condanna del maestro; così che la scienza si trasferì al di là delle Alpi. John Milton, ricordando una visita fatta, a suo dire, al "celebre Galileo, oramai vegliardo e prigioniero dell'Inquisizione" (Galileo morì nel 1642), riassunse magistralmente la situazione annotando nel 1644 che <<lo stato di servitù, in cui la scienza era stata ridotta nella loro patria, era la cagione per cui lo spirito italiano, così vivo un giorno, si era ormai spento; e da molti anni tutto ciò che si scriveva non era che adulazione e banalità>>. Non possiamo non aggiungere che abbiamo parlato della Chiesa solo perché allora essa costituiva il potere predominante: "condanne" analoghe, e frequenti, avvengono anche oggigiorno, ma ad opera degli attuali più terreni poteri che controllano le attività scientifiche.

Devono poi passare quasi due secoli prima che si riveli un altro grande fisico, Volta. Alessandro Volta genera un filone di ricerche le quali portano alle applicazioni prevalentemente tecnologiche di Antonio Pacinotti, Galileo Ferraris e Augusto Righi e, più tardi, a quelle di Guglielmo Marconi. Ma non ne deriva una vera "scuola", tanto che alla fine del 1926, quando Fermi ottiene la cattedra di fisica teorica a Roma, l'Italia certo non emerge nel mondo per la fisica.

E' solo Fermi che, ben tre secoli dopo Galileo, riesce a generare di nuovo un esteso e moderno movimento in seno alle scienze fisiche italiane. Ad esempio l'articolo di Fermi che dà avvio alla teoria delle interazioni deboli (coronata dopo cinquant'anni, nel 1983, dalle scoperte sperimentali di Carlo Rubbia [premio Nobel 1984] e collaboratori) esce nel 1933: esattamente trecento anni dopo la condanna definitiva della teoria galilieana. Le scienze fisiche italiane, nonostante i tentativi degli ultimi governi italici di distruggerle, sono ancora tra le migliori del mondo: e speriamo, ma temiamo di sbagliarci, lo restino ancora per molto. E se Ettore Majorana avesse avuto un carattere diverso, e avesse pure lui creato una sua scuola, ora la fisica italiana potrebbe trovarsi a livelli impensabili.

CAPITOLO 2

## AMICI E COLLEGHI

### Il gruppo di Roma

La rinascita della fisica italiana dovuta all'attività del gruppo di Enrico Fermi (al quale Ettore Majorana appartenne) non avrebbe avuto luogo, forse, senza l'intervento di un altro grande siciliano, Orso Mario Corbino, laureato in fisica a Palermo, professore prima a Messina e poi a Roma, dal 1918 direttore dell'Istituto Fisico di Roma. Nel 1920 era stato nominato senatore e l'anno seguente Ministro dell'Educazione Nazionale nel primo governo Mussolini. Ricorda Amaldi: <<Quando Fermi, nell'autunno 1926, iniziava a 25 anni la sua attività di professore di Fisica Teorica a Roma [...] non mancavano professori di vecchio stile, che stentavano o addirittura non volevano riconoscere il suo valore. L'intesa con O.M.Corbino era invece perfetta... Anzitutto fu chiamato, da Firenze a Roma, Franco Rasetti... L'abilità sperimentale di Rasetti completava notevolmente quella di Fermi, che a quell'epoca rivolgeva quasi tutte le sue energie alla fisica teorica. Creato così un gruppo di docenti giovani, attivi in fisica moderna e di grande valore, Corbino, Fermi e Rasetti si adoperarono per raccogliere allievi. Corbino, per esempio, rivolse un appello agli studenti di Ingegneria durante una lezione annunciando che si era aperto un periodo eccezionale per i giovani che si sentivano portati per la Fisica e disposti ad uno sforzo non comune di studio. Fu così che in via Panisperna 89A, si formò una vera e propria scuola di fisica moderna>>: ovvero, il noto gruppo dei "ragazzi di via Panisperna".

Fra gli allievi teorici ricordiamo, in ordine di ingresso in Istituto: Ettore Majorana, Gian Carlo Wick, Giulio Racah, Giovanni Gentile Jr., Ugo Fano, Bruno Ferretti e Piero Caldirola. Gli allievi nel campo sperimentale furono: Emilio Segré, Edoardo Amaldi, Bruno Pontecorvo, Eugenio Fubini, Mario Ageno e Giuseppe Cocconi.

### Gli studi di Majorana

Nel 1923 il diciassettenne Majorana (che viveva a Roma dall'età di otto o nove anni in collegio, insieme con fratelli e cugini) si era iscritto al corso di laurea in Ingegneria dell'università di Roma. Aveva come compagni il fratello Luciano, Emilio Segré, Gastone Piqué, Enrico Volterra, Giovannino Gentile e Giovanni Enriques. Gli ultimi tre erano figli, rispettivamente, del matematico Vito Volterra (uno dei pochi professori universitari --tra l'altro-- a rifiutare il giuramento al regime fascista), del filosofo Giovanni Gentile, e di Federigo Enriques, matematico ed epistemologo. Nel giugno del 1927 Corbino rivolge il suo appello agli studenti di Ingegneria: Edoardo Amaldi ed Emilio Segré ne raccolgono l'invito, ed iniziano, nel nuovo ambiente, a parlare delle doti straordinarie di Ettore; contemporaneamente, convincono Majorana ad incontrare Fermi. Il passaggio di Ettore a Fisica ha luogo all'inizio del suo quinto anno di studi universitari, dopo un colloquio con Fermi narrato da Segré, e di cui diremo. Amaldi racconta: <<Fu in quell'occasione che vidi Majorana per la prima volta. Di lontano appariva smilzo, con un'andatura timida e quasi incerta; da vicino si notavano i capelli nerissimi, la carnagione scura, le gote lievemente scavate, gli occhi vivacissimi e scintillanti: nell'insieme l'aspetto di un saraceno>>.

Più tardi si unì al gruppo Bruno Pontecorvo; così che questo risultò formato essenzialmente da Fermi, Rasetti, Majorana, Amaldi, Segré e Pontecorvo, oltre all'ottimo chimico Oscar D'Agostino.

Fermi riceverà il premio Nobel nel 1938: uno dei pochissimi premi Nobel guadagnati per ricerche svolte interamente in Italia. Del gruppo romano, solo Amaldi resterà in patria, sobbarcandosi al peso maggiore per mantenere viva nel nostro Paese la scuola creata da Fermi; e sarà pure uno dei padri dei laboratori europei di Ginevra. Gli altri emigreranno: Fermi e Segré negli Stati Uniti (ove Segré riceverà nel 1959 il suo discusso premio Nobel per la scoperta dell'antiprotone); Pontecorvo --nel settembre 1950, dopo un lungo periodo in USA, Canada, Inghilterra-- in Unione Sovietica (ove, noto universalmente come Bruno Maksimovich, diverrà presto Accademico delle Scienze dell'URSS, nonché uno dei direttori dei grandi laboratori di Dubna). Rasetti, invece, dopo avere svolto con tanto successo il ruolo di principale fisico sperimentale del gruppo romano, abbandonerà in seguito la fisica, divenendo un paleontologo di fama, e più tardi un botanico di gran valore: la abbandonerà (lui sì!), *quando* alcuni dei suoi colleghi smetteranno i panni dello scienziato per vestire quelli del tecnologo, e accetteranno di collaborare al "Progetto Manhattam" che a Los Alamos si proponeva di costruire la bomba atomica.

**Franco Rasetti**
La sua è stata una personalità così straordinaria, che vogliamo riportare qui un passo della lettera che ci scrisse da Waremme (Belgio) il 22.6.1984, all'età di quasi ottantatre anni, in risposta alla nostra consueta richiesta di "testimonianza": <<La prego di scusarmi se rispondo soltanto oggi alla Sua del 30/4 scorso, arrivata durante la mia assenza per un viaggio nell'Italia centro-meridionale a scopo botanico durato dal 6/5 al 17/6... Un precedente malessere mi aveva fatto rinunciare a un viaggio a Palermo, che aveva come scopo di fotografare le orchidacee caratteristiche della Provincia di Palermo. Come Lei saprà, da molti anni ho perso ogni interesse per la fisica, dedicandomi con molta soddisfazione e successo prima alla paleontologia del Cambriano di Canada, Stati Uniti e Sardegna, poi alla botanica delle Alpi, e per ultimo alle Orchidacee italiane... Mi dimenticavo di dirle che sarebbe vano aspettarsi da me informazioni su Majorana che non siano già note. Io ero anzitutto più anziano di lui. [Inoltre con Segré e Amaldi giocavamo a tennis, ma soprattutto eravamo compagni di alpinismo (per esempio, abbiamo fatto varie ascensioni difficili, sempre senza guida, tra cui la traversata del Cervino); mentre Majorana era estraneo ad attività sportive]. Io avevo meno relazioni con lui degli altri: se Majorana parlava con qualcuno, era piuttosto con Amaldi e Segré, suoi coetanei; o con Fermi, col quale solo poteva discutere di fisica teorica>>. E altrove, se ci è concessa una pennellata umoristica, ci scrive, quattro anni dopo: <<Ho perduto molto la memoria, e questo fatto presenta anche qualche vantaggio: cioè che rileggo con interesse e piacere anche libri che ho già letto di recente, non ricordandone più i particolari; e tra questi vi è particolarmente il Suo...>>.

## I Concorsi a Cattedre in fisica teorica

Il medesimo concorso che aveva assegnato a Fermi la cattedra romana ne attribuì una anche a Enrico Persico (a Firenze) e ad Aldo Pontremoli (a Milano), quest'ultimo perito nella spedizione Nobile col dirigibile "Italia" al Polo Nord. Persico ebbe a Firenze una influenza forte e benefica; anche i fisici fiorentini avevano da poco passato i vent'anni, erano pieni di entusiasmo, energia e idee originali. Come scrive Segrè, <<i gruppi di Firenze e Roma erano legati da sincera amicizia e spesso si scambiavano visite o seminari>>.

Passarono dieci anni prima che si aprisse un nuovo Concorso: quello del 1937 al quale, dopo alcuni anni di semi-isolamento dal gruppo di Fermi, partecipò anche Ettore Majorana. Ma di questo argomento, dai risvolti inconsueti, parleremo nel prossimo capitolo.

# CAPITOLO 3

## ETTORE PARTECIPA A UN CONCORSO

### Il Concorso a Cattedre del 1937
Dopo il concorso del 1926, in cui ottenne la cattedra Enrico Fermi, passarono altri dieci anni prima che si aprisse, nel 1937, un nuovo concorso per la fisica teorica, richiesto dall'università di Palermo per interessamento di Emilio Segré. Le vicende di questo Concorso, e specialmente i suoi antecedenti, hanno dato luogo nel 1975 ad una vivace polemica tra Leonardo Sciascia, Edoardo Amaldi, e altri (Segré; Zichichi; e chi scrive, che trovò allora ampia ospitalITà su "La Stampa" e "La Sicilia"). Ne parleremo in futuro.

Qui ci limitiamo a riprodurre i documenti certi, esistenti presso l'Archivio Centrale dello Stato. I concorrenti furono numerosi, e molti di essi di elevato valore; soprattutto quattro: Ettore Majorana, Giulio Racah (ebreo, che successivamente passerà da Firenze in Israele fondandovi la fisica teorica), GianCarlo Wick (di madre torinese e nota antifascista), e Giovanni Gentile Jr. (come sappiamo, figliolo dell'omonimo filosofo, già ministro --come si direbbe ora-- della Pubblica Istruzione), ideatore delle "parastatistiche" in meccanica quantica. La commissione giudicatrice era costituita da: Enrico Fermi (presidente), Antonio Carrelli, Orazio Lazzarino, Enrico Persico e Giovanni Polvani.

Il verbale n.1 recita: <<La commissione nominata da S.E. il Ministro dell'Educazione Nazionale... si è riunita il giorno 25 ottobre 1937-XV... Dopo esauriente scambio di idee, la Commissione si trova unanime nel riconoscere la posizione scientifica assolutamente eccezionale del Prof. Majorana Ettore che è uno dei concorrenti. E pertanto la Commissione decide di inviare una lettera e una relazione a S.E. il Ministro per prospettargli l'opportunità di nominare il Majorana professore di Fisica Teorica per alta e meritata fama in una Università del Regno, indipendentemente dal concorso chiesto dalla Università di Palermo...>>

### La relazione della Commissione del concorso del 1937
La lettera inviata lo stesso giorno a S.E. il Ministro, sulla quale il ministro Giuseppe Bottai vergò a mano la parola "Urgente", ripete il contenuto del verbale, dichiarando il Prof. Majorana Ettore avere tra i concorrenti una posizione scientifica nazionale e internazionale di tale risonanza che <<*la Commissione esita ad applicare a lui la procedura normale dei concorsi universitari*>>. Tale lettera ha un allegato, "Relazione sulla attività scientifica del Prof. Ettore Majorana". Vediamola, perché essa ci dà sintetiche informazioni sull'opera scientifica del Nostro: <<Prof. Majorana

Ettore: si è laureato in Fisica a Roma nel 1929. Fin dall'inizio della sua carriera scientifica ha dimostrato una profondità di pensiero ed una genialità di concezione da attirare su di lui l'attenzione degli studiosi di Fisica Teorica di tutto il mondo. Senza elencarne i lavori, tutti notevolissimi per l'originalità dei metodi impiegati e per l'importanza dei risultati raggiunti, ci si limita qui alle seguenti segnalazioni. Nelle teorie nucleari moderne il contributo portato da questo ricercatore con la introduzione delle forze dette "Forze di Majorana" è universalmente riconosciuto, tra i più fondamentali, come quello che permette di comprendere teoricamente le ragioni della stabilità dei nuclei. I lavori del Majorana servono oggi di base alle più importanti ricerche in questo campo. Nell'atomistica spetta al Majorana il merito di aver risolto, con semplici ed eleganti considerazioni di simmetria, alcune tra le più intricate questioni sulla struttura degli spettri. In un recente lavoro infine ha escogitato un brillante metodo che permette di trattare in modo simmetrico l'elettrone positivo e negativo, eliminando finalmente la necessità di ricorrere all'ipotesi estremamente artificiosa ed insoddisfacente di una carica elettrica infinitamente grande, diffusa in tutto lo spazio, questione che era stata invano affrontata da molti altri studiosi>>.

Uno dei lavori più importanti di Ettore, quello in cui introduce la sua "equazione a infinite componenti" (di cui diremo), non è menzionato: ancora non era stato capito. È interessante notare, però, che viene dato opportuno rilievo alla sua teoria simmetrica per l'elettrone e l'anti-elettrone (oggi in auge, per la sua applicazione a neutrini e anti-neutrini); e a causa della capacità di eliminare l'ipotesi cosiddetta "del mare di Dirac" [P.A.M. Dirac, premio Nobel 1933]: ipotesi che viene giustamente definita artificiosa e insoddisfacente.

Una volta attribuita la cattedra a Ettore "fuori concorso" --applicando una legge che era stata usata per dare una cattedra universitaria a Guglielmo Marconi [premio Nobel 1909]--, la commissione riprendeva i suoi lavori giungendo all'unanimità alla formazione della terna vincente: Gian Carlo Wick (che andò a Palermo); Giulio Racah (a Pisa); e Giovannino Gentile (a Milano). Quest'ultimo, grande amico di Ettore, scomparirà prematuramente nel 1942; non prima di avere pubblicato postumo un articolo divulgativo di Majorana sulle similarità tra le leggi statistiche della nuova fisica quantistica e quelle delle scienze sociali (da noi riportato nelle Appendici, in forma opportunamente ridotta). Alcune belle e interessanti lettere di Ettore a Giovannino Gentile ci pervennero per il tramite della famiglia Gentile di Torino, di Leonardo Sciascia e di Luciano Canfora [ogni lettera, così come ogni altro documento, è stata da noi pubblicata nel volume dello scrivente su "Il Caso Majorana: Epistolario, Testimonianze, Documenti"]: in una di esse, del 1934, Ettore esprime all'amico sentimenti analoghi a quelli esposti nell'articolo postumo, comunicandogli la propria soddisfazione per il fatto che la Fisica aveva cessato di "servire da giustificazione per il volgare materialismo". L'articolo divulgativo era stato, però, cestinato dal Majorana; e si è salvato solo per intervento esterno dei fratelli di Ettore.

**Gli atti ufficiali**

Il 2 novembre 1937 il ministro Bottai emette il decreto di nomina a professore ordinario di Ettore Majorana; il 4 dicembre il decreto è già registrato alla Corte dei Conti; e quindi la nomina viene partecipata dal Ministero a Ettore, presso la sua abitazione di viale Regina Margherita 37, in Roma. Lo vedremo più avanti; ora chiediamoci quanto segue.

**Un dubbio**

Come mai una persona lontana dagli interessi terreni quanto Majorana partecipò a un concorso a cattedre? Per avere degli studenti, sospettavamo. Infatti, chi pensa, fa, scopre cose grandi, non può disprezzare i propri risultati e il proprio lavoro: e immaginavamo un Majorana desideroso di trasmettere a degli allievi ciò che la sua mente prodigiosa andava scoprendo delle leggi del Creato. La conferma la si è avuta pochi anni or sono, quando il mio collaboratore Salvatore Esposito di Napoli ha scoperto, insieme con un collega romano, che, anche durante gli anni del presunto isolamento di Ettore Majorana dal mondo e dai colleghi, ovvero durante gli anni accademici 1933-34, 1934-35 e 1935-36, Ettore presentò domanda al fine di potere tenere dei corsi universitari ``liberi'' (e gratuiti) presso l'Istituto di Fisica romano di Via Panisperna: cosa che gli era permessa, avendo egli già conseguito quella che si chiamava Libera Docenza. Il direttore, Orso Mario Corbino, fece approvare tali richieste; ma non risulta che Majorana abbia tenuto quei corsi: probabilmente perché gli studenti che si rendevano conto dell'importanza degli argomenti relativi erano troppo pochi, in quegli anni…

# CAPITOLO 4

## IL NONNO E GLI ZII DEPUTATI

*Una famiglia di deputati e studiosi*

Fondatore della famiglia di Ettore può essere considerato il nonno, Salvatore Majorana Calatabiano, nato a Militello Val di Catania nel 1825. Da semplice agricoltore, nel 1865 diventa professore ordinario all'università di Messina (e poco dopo a Catania), e l'anno successivo viene eletto deputato al parlamento. Nel primo ministero della Sinistra, il capo del governo, Depretis, gli affida il portafoglio di Agricoltura, Industria e Commercio; e, dopo la crisi del 1877, torna al suo posto di ministro nel terzo governo Depretis. Ci può interessare qui una citazione. Convinto che le leggi economiche siano leggi naturali di indole matematica, scrive: <<è lo sprezzo dei dettami scientifici --che in conclusione dovrebbero essere nel campo delle cose legislative quello che nelle applicazioni tecniche sono i teoremi della fisica e del calcolo,--codesto sprezzo, codesto divorzio tra il pensiero e la pratica, tra la scienza e l'arte sociale, la causa potente del disagio in cui la cosa pubblica si trova>>.

Sposatosi in seconde nozze con Rosa Campisi, Salvatore ne aveva avuto sette figli: Giuseppe, Angelo, Quirino, Dante, Fabio Massimo (il padre di Ettore), Elvira ed Emilia. Tre dei figli arrivano ad essere deputati, nonché rettori dell'università di Catania (sì, tre fratelli rettori della stessa università: rispettivamente negli anni 1895-1898, 1911-1919 e 1944-1947).

Quirino si laurea a 19 anni in Ingegneria e a 21 in Scienze fisiche e matematiche, e diviene poi presidente della Società Italiana di Fisica (sarà anche membro della commissione che diede la cattedra a Fermi). Famosi per la loro eccezionale precisione sono i suoi esperimenti su un possibile *assorbimento* della gravitazione nella materia. Elvira ed Emilia compiono la loro educazione ed istruzione a Roma, e quivi si sposano.

Anche Fabio, il padre di Ettore (n. a Catania nel 1875, m. a Roma nel 1934), si laurea giovanissimo --a 19 anni-- in Ingegneria, e poi in Scienze fisiche e matematiche. Sarà lui a educare culturalmente e scolasticamente il piccolo Ettore fino agli otto o nove anni. Ettore resterà sempre molto attaccato al padre; e senza dubbio ne sentirà profondamente la dipartita, nel 1934. L'ing. Fabio Majorana fonda a Catania la prima impresa telefonica, e più tardi diviene ispettore generale del Ministero delle comunicazioni. Scrive Edoardo Amaldi: <<Dal matrimonio dell'Ing. Fabio con la Sig.ra Dorina Corso (n. a Catania nel 1876, m. a Roma nel 1966), anch'essa di famiglia catanese, nacquero cinque figli: Rosina, sposata più tardi con Werner Schultze; Salvatore, dottore in legge e studioso di filosofia; Luciano, ingegnere civile, specializzato in costruzioni aeronautiche ma che poi si dedicò alla progettazione e costruzione di strumenti per l'astronomia ottica; Ettore (n. a Catania il 5 agosto 1906); e, quinta e ultima, Maria, musicista... >>.

Ma torniamo un momento agli zii, rettori e deputati, di Ettore. Dovendo scegliere, accenniamo solo ad Angelo: cordiale, di bell'aspetto, ed oratore seducente. Brilla precocissimo, ma presto si spegne; e la parabola della sua vita lascia pensosi, quando la si paragoni --per quanto ne sappiamo-- a quella di Ettore. Angelo nasce a Catania, secondo dei sette figli di Salvatore, nel 1865. Maturo a 12 anni, all'età di 16 è dottore in legge a Roma. Conseguita la libera docenza a 17 anni, si presenta a tre concorsi nel 1886, per le università di Catania, Messina e Pavia: e, a venti anni, li vince tutti e tre. Passa così titolare a Catania, ove diviene rettore a 29 anni, cominciando nel contempo a dedicarsi altresì alla politica attiva. Di coloritura liberale (<<lo stato deve essere garanzia di tutte le libertà e della libertà di tutti>>), il primo ministro Giolitti gli affida nel 1904 il dicastero delle Finanze e, pochi anni dopo, quello del Tesoro. Ma presto il suo organismo si consuma, esausto per l'imponente attività sostenuta. Si spegne a Catania a soli 44 anni.

*Ettore Majorana: Il ricordo dei colleghi*
L'ammirazione dei colleghi per gli aspetti scientifici ed umani di Ettore Majorana è ben rappresentata dalle seguenti poche parole che ci scrisse nel 1984 l'illustre fisico Gilberto Bernardini dalla Scuola Normale di Pisa: <<...Ricordo che io, con Ettore, evitavo di parlare di fisica, perché quello che avrei potuto dirgli sarebbe stato per lui insignificante... Vorrei però prescindere dall'eccezionale ingegno di Ettore come fisico per rievocarne la complessa spiritualità emotiva ed umana, tanto più estesa ed illuminata di quella sulla quale in tanti hanno fantasticato...>>

Ettore era persona sensibilissima e introversa, ma profondamente buona. La sua ritrosia e timidezza e la sua difficoltà di contatto umano --reso ancor più difficile dalla sua stessa intelligenza--, non gli impedivano di essere sinceramente affettuoso. E la sua critica severa si addolciva quando il giudizio riguardava gli amici.

Molti sanno infatti dell'eccezionale spirito critico, e autocritico, di Ettore; ma pochi sanno che, almeno fino al 1933 (anno in cui Ettore trascorse vari mesi a Lipsia, presso Werner Heisenberg) Ettore era di carattere allegro. La sorella Maria ne ricordava soprattutto le barzellette, le risate, il gioco alla palla fatto nel corridoio di casa in Via Etnea 251. Da ragazzo, con amici, si mise al volante della macchina di famiglia --una delle prime auto--, ovviamente senza saper guidare, e altrettanto ovviamente finì contro un muro, serbandone ricordo con una lunga cicatrice su una mano.

Ettore era poi ricchissimo di *humour*, cosa abbondantemente confermata dai tanti episodi aneddòtici e dal suo stesso epistolario. Un esempio? Da Lipsia, nel 1933, scrive alla madre: <<...Io mi trovo benissimo... Nevica spesso dolcemente... L'Istituto di Fisica Teorica con molti altri affini è posto in posizione *ridente*, un po' fuori di mano, tra il cimitero e il manicomio... All'Istituto mi hanno accolto molto cordialmente: ho avuto una lunga conversazione con Heisenberg, che è una persona straordinariamente cortese e simpatica, e ama le mie chiacchiere… Sto già scrivendo alcuni articoli in tedesco… La situazione politica interna appare permanentemente catastrofica, ma non mi sembra che interessi molto la gente… Ho fatto un viaggio eccellente.  Dal Brennero a Monaco ho avuto soltanto la compagnia di un gentiluomo napoletano, ubriaco fradicio.  Era diretto a Brema per vendere pomodori, e mi ha onorato di tutta la sua confidenza perché ha ammesso a priori che io mi recassi a Lipsia per lo stesso scopo…>>.

# CAPITOLO 5

## ETTORE MAJORANA: LA SCOMPARSA

*Premessa:*

Dovremmo ora passare ad esaminare un po' più a fondo la personalità umana, e soprattutto scientifica, di Ettore Majorana. Ma il persistente mito della sua "scomparsa" ci obbliga a questo punto ad affrontare tale delicato e personalissimo tema. Prima ricordiamo, però, alcune frasi conclusive scritteci da Gianni Sansoni, dopo avere esaminato la calligrafia del Nostro: <<…Come vedi, non possiamo chinarci su di una persona senza iniziare un discorso che non può che essere molto lungo… Posso però dire questo: che Majorana doveva essere anche persona mite e buona, bisognosa d'affetto più che mai, e penso che migliore elogio non gli si possa fare che avvicinandoci alle sue vicende con rispetto e comprensione>>.

Ci piace anche premettere un ricordo proferito da Maria, indimenticabile sorella di Ettore (ora scomparsa) così come ce lo ricordiamo, inciso indelebilmente nella nostra memoria: <<Era schivo e timido, di spirito arguto; con un vivo senso dell'umorismo, e una enorme sensibilità umana... Io ero la sua sorellina più piccola, e mi voleva molto bene. Era così gentile che mi faceva anche i compiti di matematica... Ho molti ricordi d'infanzia. D'autunno andavamo in villeggiatura sull'Etna. Nelle notti senza luna Ettore mi indicava il cielo, le stelle, i pianeti: ogni volta era una piccola lezione di astronomia. Le sue parole mi tornano alla mente ancora oggi, quando alzo lo sguardo verso il cielo stellato... Mi piace ricordarlo così, mentre mi invita a guardare il cielo e mi insegna a chiamare per nome le stelle>>.

*La scomparsa:*

Il venerdì 25 marzo 1838 (dopo pochi mesi di insegnamento della fisica teorica presso l'università di Napoli, ove era stato nominato professore ordinario –fuori concorso— per chiara fama), Majorana spedisce al direttore, Carrelli, della Fisica di Napoli, scrivendo con calligrafia tranquillissima, la seguente lettera: << *Caro Carrelli, Ho preso una decisione che era oramai inevitabile. Non c'è in essa un solo granello di egoismo. Ma mi rendo conto delle noie che la mia improvvisa scomparsa potrà procurare a te e agli studenti...dei quali tutti conserverò un caro ricordo almeno fino alle 11 di questa sera, e possibilmente anche dopo*>>. Esce di casa, ovvero dal suo albergo, il "Bologna" di Napoli, alle 17; per imbarcarsi sulla nave Napoli-Palermo delle 10:30 della sera: il "Postale", cioè, della compagnia Tirrenia. Sul tavolo della sua stanza ha lasciato una lettera *Alla mia famiglia,* che riteniamo non necessario riprodurre (diversamente da quanto fatto da molti, illecitamente e per motivi egocentrici o tornaconto personale): le persone famose devono rinunciare a molta della propria privacy, è vero, ma lo spirito di discrezione non può mai fare difetto. Rischiando di essere al limitare della vita personale di Ettore, è bene gli ultimi tratti del cammino percorrerli con piedi leggeri: coi passi dell'intuizione, dell'immaginazione, della sensibilità di ciascuno di noi. Uscito dall'hotel, cosa avrà pensato, e fatto Ettore in quelle cinque ore…?  Forse era dal periodo della sua vita indipendente in Lipsia, presso l'istituto del fisico-filosofo Werner Heisenberg -- ovvero da cinque anni--, che pensava alla necessità di questa sua scomparsa: trattenuto forse dalla preoccupazione di causare dolore alla famiglia.

Il 26 marzo, sabato, da Palermo manda un telegramma a Carrelli perchè non dia peso alla sua precedente lettera, e un telegramma al suo hotel di Napoli, e quindi, su carta (eccezionalmente) intestata dell'Hotel Sole di Palermo, un espresso a Carrelli: espresso che viene consegnato la domenica  mattina al destinatario: <<...*Il mare mi ha rifiutato, e ritornerò... Ho però intenzione di rinunciare all'insegnamento... Non mi prendere per una ragazza ibseniana, perchè il caso è differente...*>>. Questo è l'ultimo documento scritto da Majorana che si possegga: l'ultimo manoscritto del suo progetto, della sua "messinscena"?

Ma durante il viaggio di ritorno da Palermo a Napoli, o subito prima, o subito dopo, Ettore fa perdere le proprie tracce.  Si uccise?  Non lo crediamo (si era fatto mandare la parte del suo conto in banca un mese prima; aveva ritirato per la prima volta lo stipendio dei primi mesi di docenza universitaria; aveva preso con sé il passaporto; e così via).  Andò in un convento?  Può darsi: ma esistono solo due testimonianze di suoi probabili approcci con Conventi del napoletano, onde fare esperimento di vita religiosa: tentativi da lui poi abbandonati di fronte alle esigenze burocratiche oppostegli.  Certamente, alla luce delle sue ultime lettere –scritte, ripetiamolo, con calligrafia ordinata e tranquilla— si devono scartare del tutto le ipotesi cervellotiche dell'omicidio o del rapimento.  I documenti più seri ci parlarono di una sua fuga in Argentina: fuga forse dai propri "pupi" pirandelliani di membro regolare di un gruppo di ricerca in fisica (quando in fisica teorica superava di dieci lunghezze tutti gli altri), e di bravo figlio di una famiglia del Sud Italia –così pensano anche molti dei familiari attuali-- con una madre di carattere dominante... Si tratta di documenti seri, dicevamo: ma non conclusivi!; anche se vi prestarono attenzione fisici illustri come Yuval  Ne'eman, Tullio Regge e Remo Ruffini,  e scrittori (nelle sue lettere private a noi  spedite) come Leonardo Sciascia.

Ettore teneva sul comodino libri di Shakespeare, di Shopenauer,  di Pirandello...
E, ne *Il fu Mattia Pascal,*  già nel 1904 Luigi Pirandello aveva fatto dichiarare al suo protagonista:  <<Chissà quanti sono come me, nelle mie stesse condizioni, fratelli miei.  Si lascia il cappello e la giacca,  con una lettera in tasca, sul parapetto d'un ponte, su un fiume; e poi,  invece di buttarsi giù, si va via tranquillamente: in America o altrove>>.  Più ancora, Vitangelo Mostarda, in  *Uno, Nessuno, Centomila*, ancora di Pirandello (1926), dichiara: <<A vedermi addosso gli occhi di tutta quella gente, mi pareva di sottostare a una orribile sopraffazione pensando che tutti quegli occhi mi attribuivano una immagine che non era certo quella che io mi conoscevo, ma un'altra che io non potevo né conoscere né impedire>>... Il critico letterario Croci commenta: <<Vitangelo Moscarda si trova impegnato in un disperato esperimento: quello di ricostruirsi una esistenza svincolata dai condizionamenti imposti dalla natura e dalle convenzioni, e di affermare la propria personalità autentica mediante un atto di libera scelta>> (il quale non esclude, a priori, il convento: chi sapeva scorgere nella natura l'entusiasmante perfezione e l'eccelsa bellezza e profondità delle sue leggi ben avrebbe potuto cercare alfine rifugio presso la Mente suprema che tale natura gli rivelava).

Dopo un nostro recente intervento al "Kiwanis", il Procuratore generale della repubblica di Catania ha osservato che si può ritenere che Ettore fosse un genio non solo nella fisica, ma in tutto: anche nel resto... Difatti, come ricorderemo, il grande fisico Gilberto Bernardini ci aveva analogamente raccomandato "di prescindere dall'eccezionale ingenio di Ettore come fisico, per accentuare anche quanto possa rievocare la sia complessa spiritualità umana, tanto estesa ed illuminata...". E una nostra critica, l'italiana Aurora Bernardini di San Paolo del Brasile, in completa sintonia, ebbe a scriverci: <<...stando a quanto si sa di Majorana non rimane che credere che in lui la genialità abbia anticipato la scoperta della sua verità. O della verità *tout court*, che Ivan Ilich di Tolstoj scopre solo prima di morire. Quali sono i momenti veramente vivi della vita? Ognuno ha la sua risposta, quasi sempre in ritardo. Majorana l'avrebbe avuta prima. Sarebbe molto utile per l'odierna umanità il suo legato in proposito. Forse ancora più utile --*honni soit...*-- che il suo legato in quanto fisico>>.

# CAPITOLO 6

## IL GENIO CHE SCOMPARE

**Il genio.**
Nella lettera scritta il 27 luglio 1938 a Mussolini, e che già abbiamo incontrato, Enrico Fermi --dopo avere dichiarato che fra tutti gli scienziati il Majorana era quello che più al mondo l'avesse colpito-- aggiunge: <<Capace nello stesso tempo di svolgere ardite ipotesi e di criticare l'opera sua e degli altri; calcolatore espertissimo e matematico profondo che mai per altro perde di vista sotto il velo delle cifre e degli algoritmi l'essenza reale del problema fisico, Ettore Majorana ha al massimo grado quel raro complesso di attitudini che formano il fisico teorico di gran classe. Ed invero, nei pochi anni in cui si è svolta fino ad ora la sua attività, egli ha saputo imporsi all'attenzione degli studiosi di tutto il mondo, che hanno riconosciuto in lui uno dei forti ingegni del nostro tempo. E le successive notizie della sua scomparsa hanno costernato quanti vedono in lui chi potrà ancora molto aggiungere al prestigio della Scienza Italiana>>. Un'altra volta, come già abbiamo visto, Fermi aveva paragonato Ettore a Galileo e Newton; considerandolo a tutti, e a se stesso, superiore nel campo della fisica teorica. Tanto che (come ci ha raccontato Piero Caldirola) una volta Bruno Pontecorvo rimproverò Fermi di "umiliarsi" troppo di fronte ad Ettore.

La fama di Ettore Majorana è cresciuta e cresce col tempo, lo si è detto, anche tra i colleghi: così come sempre avviene quando essa corrisponde a verità. Da varie decine di anni una elevata percentuale di pubblicazioni scientifiche nel mondo (in alcuni settori della fisica) suole contenere il suo nome anche nel titolo. Lui non era certo inferiore a un Wigner (premio Nobel 1963) o a un Weyl: i quali, per il loro rigore fisico-matematico, erano forse gli unici fisici che suscitassero da parte di Ettore una approvazione senza riserve. La sua acutezza lo portava infatti a vedere molto al di là dei colleghi: ad essere, come sappiamo, un grande e lungimirante pioniere.

Perfino i dettagli del primo incontro di Majorana con Fermi, avvenuto allorché, alla fine del 1927, Ettore cominciò concretamente a pensare di lasciare l'Ingegneria per la Fisica, ci illuminano circa alcuni aspetti scientifici (e anche non scientifici) del Majorana. Essi sono noti da quando li ha narrati, ad esempio, Emilio Segré: <<Il primo lavoro importante scritto da Fermi a Roma [su alcune proprietà statistiche dell'atomo]… è oggi noto come metodo di Thomas-Fermi… Quando Fermi trovò che per procedere gli occorreva la soluzione di un'equazione differenziale non lineare, caratterizzata da condizioni al contorno insolite, con la sua abituale energia in una settimana di assiduo lavoro calcolò la soluzione con una piccola calcolatrice a mano. Majorana decise che probabilmente la soluzione *numerica* di Fermi era sbagliata e che sarebbe stato meglio verificarla. Andò a casa, trasformò l'equazione originale di Fermi in una equazione del tipo di Riccati e la risolse senza l'aiuto di nessuna calcolatrice, servendosi della sua straordinarie attitudini matematiche… Quando tornò in Istituto confrontò con aria scettica il pezzetto di carta, su cui aveva riportato i dati ottenuti, col quaderno di Fermi, e quando trovò che i risultati coincidevano esattamente non poté nascondere la sua meraviglia>>. In realtà Ettore trasformò l'equazione di Thomas-Fermi in una equazione di Abel (e non di Riccati), come ci hanno detto i manoscritti da lui lasciati inediti. Ma, non contento, inventò un secondo metodo, tuttora sconosciuto, che gli permise di risolvere "matematicamente" (cioè *analiticamente,* e non più solo numericamente) l'equazione di Thomas-Fermi.

Il lavoro eseguito da Ettore in poche ore ha richiesto, per essere compreso a fondo, *due mesi* di intensa applicazione di un provetto fisico teorico quale il nostro collaboratore Esposito. Fermi non aveva esagerato nel ritenere il Majorana un genio al pari di Galileo e Newton…

Dovremo, quindi, prestare attenzione soprattutto alla genialità di Ettore Majorana, davanti alla quale ogni diversa considerazione perde di importanza. E ciò continueremo a fare. Ma il mito della "scomparsa" –che ha contribuito, si diceva, a dare al Majorana null'altro che la notorietà che gli spetta-- ci impone e ancor più ci imporrà di indugiare ancora sugli ultimi giorni della vita pubblica del Majorana.

**La scomparsa.**

Dopo il 26 marzo del 1938, data dell'ultima lettera nota di Ettore, inviata da Palermo a Carrelli, direttore dell'Istituto di Fisica di Napoli, il Majorana ha lasciato ulteriori sue tracce? Forse sì… Anzitutto, la sua infermiera, che bene lo conosceva, dichiara alla Polizia di averlo visto ai primi di aprile in Napoli, tra Palazzo Reale e Galleria, "mentre veniva su da Santa Lucia".

I fratelli di Ettore, poi, raccomandati dal grande filosofo Giovanni Gentile, vengono ricevuti il 18 di aprile a Roma dal Capo della Polizia, senatore Arturo Bocchini; il quale comunque, sollecitato per telefono dal Questore di Napoli, aveva già inviato il 31 di marzo il seguente cablogramma a tutti i Questori del Regno: <<Pregansi ricerche ai soli fini rintraccio, senza far nulla trapelare all'interessato, del professore fisica R.Università Napoli, Ettore Majorana... Alt Connotati: Altezza 1,68; Viso lungo; Occhi scuri et grandi; Capelli neri; Bruno; Veste soprabito grigio-ferro; Cappello marrone scuro Alt Caso rintraccio telegrafare urgenza Ministero segnalando eventuale di lui spostamento et località ove dirigesi Alt Capo Polizia>>. Questo telegramma viene trasmesso dall'Ufficio Cifra il 31 di marzo col numero 10639 [ogni documento di questo tipo è stato rinvenuto presso l'Archivio Generale dello Stato].

L'unico risultato delle indagini poliziesche è quello riportato nella comunicazione n.87966 inviata dallo stesso Questore di Napoli al Rettore dell'università partenopea in data 29 aprile: <<Riservatissima -- …A richiesta del Dott. Salvatore, fratello del Prof. Majorana Ettore, furono iniziate e poi intensificate le ricerche, ma finora con esito negativo. E` emerso soltanto che lo scomparso, pare il 12 corrente, si presentava al Convento di S. Pasquale di Portici per essere ammesso in quell'ordine religioso, ma, non essendo stata accolta la sua richiesta, si allontanò per ignota destinazione…>> [Serie "Polizia Politica: Personali", busta n.780; fascicolo "Ettore Majorana"]. Purtroppo la data del 12 Aprile non viene segnalata come sicura. Ad ogni modo, questa informazione si aggiunge alla testimonianza, più nota, del padre gesuita De Francesco, il quale, nelle foto mostrategli dalla famiglia (in particolare dal fratello Dott. Luciano), riconosce il giovane distinto che –negli ultimi giorni di marzo o ai primi di aprile-- si era presentato al Superiore della Chiesa detta del Gesù Nuovo, in Napoli, chiedendo di essere ospitato nell'annesso convento per fare esperimento di vita religiosa; il giovane, di fronte alle difficoltà burocratiche frappostegli, aveva ringraziato, si era scusato, e se n'era andato.

Ettore si ritirò quindi in un convento? Può darsi; lo si è già detto: chi sapeva scorgere nella natura l'entusiasmante perfezione e l'eccelsa bellezza e profondità delle sue leggi ben avrebbe potuto cercare alfine rifugio presso la Mente suprema che tale natura gli rivelava.

Ma le visite ai conventi potrebbero fare parte di un piano organizzato dal Majorana per sviare ogni futuro tentativo di ricerca. I documenti apparentemente più seri che abbiamo rintracciato sembrano infatti confermare –come riteneva sua madre- - una fuga in luogo appartato: lontano dai condizionamenti pesanti della sua vita italiana; fuga resa *forse* "inevitabile" da necessità imprescindibili richieste dall'equilibrio di un animo tanto sensibile e geniale. Il direttore, Carlos Ribera, dell'Istituto di fisica della università cattolica di Santiago del Cile dichiarò infatti di avere avuto a Buenos Aires, due volte, prove indirette della presenza del Majorana in Argentina. E la sua dichiarazione ci giunse dapprima attraverso Remo Ruffini, venendo poi confermata dal grande fisico israeliano Yuval Ne'eman (che era stato generale dell'esercito, nonché capo dei servizi segreti di Israele durante la Guerra dei Sei Giorni, prima di decidersi a dedicare la propria intelligenza alla ricerca scientifica piuttosto che alla vita militare). Ne'eman diede a priori un certo credito a tale segnalazione. E il notissimo fisico italiano Tullio Regge, dovendo visitare Santiago, volle parlare con Ribera: e ci mandò una dettagliata relazione circa questa sua "intervista", testimoniando che il Professor Ribera (già allievo di Erwin Schroedinger) appariva persona degna di fede. Inoltre, la signora Blanca de Mora, vedova del premio Nobel guatemalteco Asturias, confermò che a Rosario o Santa Fé viveva, o era vissuto almeno fino agli inizi degli anni Sessanta, un certo Ettore Majorana, fisico, che scendeva a Buenos Aires per frequentarvi il salotto culturale delle sorelle Manzoni –discendenti del grande romanziere--, essendo egli amico di una di esse: Eleonora Cometti-Manzoni, matematica.  Perfino Sciascia, pur nel suo rigore intellettuale, in una propria lettera a noi indirizzata  considerò  degna di considerazione  l'ipotesi "Argentina"... Ma i documenti non dicono tutto, e non sono pertanto conclusivi. Così che, ancora una volta, la conclusione, la decisione finale non può che essere lasciata alla sensibilità e alla coscienza di ogni singolo lettore.

# CAPITOLO 7

## ANATOMIA DI UNA SCOMPARSA

### *Prima:*

Parte essenziale dell'epistolario, da noi riscoperto nel 1972 insieme con Maria Majorana, sono proprio le missive inviate da Ettore nel 1938, da Napoli, nei mesi immediatamente precedenti la sua misteriosa scomparsa. L'esame di queste mostra un salto improvviso tra il tono delle ultimissime tre lettere di Majorana e quello di tutte le lettere anteriori. Alla famiglia, ad esempio, Ettore era solito inviare scritti equilibrati (forse"controllati"), esplicativi, ricchi di umorismo, affettuosi e lunghi. Il 23 febbraio, un mese prima di sparire, racconta alla madre, dal suo albergo, il"Bologna", di Napoli: <<Oggi mi daranno una stanza su via Depretis, da cui potrò vedere *fra tre mesi* il passaggio di Hitler! Siete guarite dai vostri due piccoli raffreddori? Verrò forse dopo Carnevale. Saluti affettuosi - Ettore>>. E il 12 gennaio nel ringraziare il Ministro per "l'alta distinzione"concessagli con la nomina ad ordinario fuori concorso , aveva scritto: "tengo ad affermare che darò ogni mia energia alla scuola e alla scienza italiane".

Perciò, quando il 22 gennaio chiede che il fratello ing. Luciano gli mandi "la sua parte del conto in banca", c'è da credere che, in quel momento, pensasse alla propria sistemazione fissa in Napoli. Questa intenzione, di avere un posto dove vivere per i fatti suoi, trapela ancora da ciò che scrive al fratello dr. Salvatore una settimana prima della scomparsa: <<Napoli, 19 –3-1938 - XVI. Caro Turillo,… Vedrò se è possibile avere il libretto della Mutua per la mamma, ma non vedo come si possa affermare la convivenza, perché io ho l'obbligo di prendere la residenza a Napoli: anzi l'ho già presa provvisoriamente qui in albergo, alias via Depretis 72>>. E ci sembra che l'agire di Ettore non
sia solo un ossequio all'obbligo di risiedere nella stessa città in cui esercita l'insegnamento. Ma quello stesso sabato 19 di marzo Ettore, portato a un termine il proprio interiore travaglio, aveva probabilmente già preso la sua "ormai inevitabile" decisione. Non vi era in essa "un solo granello di egoismo": come dire che per lungo tempo, forse per anni, si era chiesto in cuor suo se poteva moralmente prendere questa decisione, o se essa gli era proibita perché dettata almeno in parte da esigenze egocentriche. Forse volgeva tali pensieri nel suo animo, a tratti, fin dal 1934, dopo il suo rientro da Lipsia; fino a convincersi della necessità di una decisione che, come tale (necessaria), era oramai purificata da ogni grano di egoismo. Ed Ettore si accinge a realizzare il suo meditato e sofferto progetto, a dare inizio alla sua costruzione (le parole che normalmente si usano in questi casi, come "messinscena", non si prestano al caso di Ettore), già, probabilmente, questo sabato. Invia infatti un telegramma a Roma col quale disdice il suo arrivo per trascorrere a casa --come faceva di consueto– la domenica. E quindi scrive al fratello maggiore: <<Per ora non vengo perché lunedì ha alcune faccende da sbrigare… Vi mando un telegramma perché non mi aspettiate stasera, ma verrò certamente sabato prossimo>>.

Poi una settimana di silenzio epistolare. E il "sabato prossimo" sarà quello (26 marzo) dell'ultima sua lettera: da Palermo, a Carrelli.

Venerdì 25 riprende in mano la penna. Nella lettera (la prima) a Carrelli, si rende conto –dice, e già l'abbiamo visto-- *delle noie che la mia improvvisa scomparsa potrà procurare a te e agli studenti.* <<Anche la parola *scomparsa*, in luogo di morte o di fine, crediamo che sia stata usata perché venisse intesa come eufemismo, mentre non lo era >>: questo l'ha detto Sciascia. Ma nella stessa lettera, nel chiudere, aggiunge:… *dei quali tutti_conserverò un caro ricordo almeno fino alle undici di questa sera, e possibilmente anche dopo.* Vuole far credere che le undici siano l'ora del pensato suicidio; esse sono invece, banalmente, l'ora della partenza: la nave era prevista partire alle 10 e 30' di sera; approssimando, o calcolando un ragionevole ritardo, le 22 e 30' diventano le 23. Infine, con le ultime parole ("e possibilmente anche dopo") vuole far credere a un estremo omaggio alle speranze della religione, mentre il loro significato, ancora una volta, è quello letterale: e questo lo aggiungiamo noi. Ma si tratta del medesimo "gioco al limite dell'ambiguità" che la sensibilità di Sciascia ha avvertito.

***Durante :***

Citare Pirandello nel caso Majorana non è certo cosa nuova. Ma senza forzare la realtà possiamo immaginare che davvero, su quello stesso comodino ove lascia la sua lettera per la famiglia, Ettore tenga Schopenhauer e Shakespeare; e Pirandello. Pure lui siciliano: gloria, insieme ad Empedocle, di Agrigento; così come Bellini, Verga e --ora– Majorana sono glorie di Catania. Ancora non conosciamo il perché della decisione di Majorana; ma le carte ce ne suggeriscono il come. Quante volte avrà letto --e, cosa più suggestiva, avrà visto a teatro– *Il fu Mattia Pascal* ?: È facile scomparire, dice il protagonista; rileggiamo le parole di Pirandello: "Si lascia il cappello e la giacca, con una lettera in tasca, sul parapetto di un ponte, su un fiume; e poi invece di buttarsi giù, si va via tranquillamente: in America o altrove". E forse Ettore esegue esattamente: sa che le cose più banali sono le meno credute. Solo che in tasca si mette passaporto e soldi (quei soldi che possiamo calcolare equivalere ad almeno 20 mila euro di oggi); e la lettera la lascia sul comodino.

Così Ettore se ne va. Con una nave; o con la *prima* delle navi del suo "progetto". In tutti gli ultimi anni, il suo noto, risvegliato amore per le navi era soltanto interesse matematico per la strategia navale, e attenzione ingegneristica ai caratteri costruttivi delle navi, o era piuttosto il sintomo esteriore di un desiderio ancora inconfessato, ma già presente, di fuggirsene lontano, oltre mare? A quel tempo chi emigrava sognava l'Argentina. Lo stesso Mattia Pascal, subito dopo aver nominato l' *America o altrove,* precisa i suoi pensieri con un nome: Buenos Aires. Ettore, così, se ne va via. Ma non *tranquillamente.*

Durante quella notte in nave, tra Napoli e Palermo, la sua mente e il suo cuore non hanno riposo; anche se riesce, ci dicono, a dormire. La polizia, i colleghi, gli amici lo crederanno morto, e non lo cercheranno: proprio come lui voleva; lo scopo che si era ripromesso con le sue lettere dalla calligrafia, come sempre, composta e ordinata. "Preordinata", dice Sciascia. Ma ancora una volta pensa: e la famiglia? la madre? intenderanno, invece, i familiari che lui ha lasciato loro una speranza? La sua decisione risponde ad esigenze oggettive; appartiene, quindi, al mondo delle cose necessarie, giuste, etiche. Ma non riceveranno i familiari un dolore troppo acerbo? I dubbi di sempre riprendono il sopravvento; anche in lui che –quando non aveva a che fare coi sentimenti umani, bensì con le serene, imperturbabili e alte cose della natura– sapeva edificare architetture di pensiero vertiginose ma stabili; sapeva "calcolare" ogni armonico rapporto con maestria ineguagliata.

E appena sbarcato a Palermo invia il telegramma urgente che conosciamo; il quale giunge nelle mani del direttore, Carrelli, dell'Istituto fisico di Napoli quella stessa mattina alle ore 11. Ettore sa che Carrelli, come tutti, ha pensato al suicidio; e nella lettera che fa seguire dice pertanto: *il mare mi ha rifiutato,* non senza una nota di amara autoironia. Forse per un poco pensa davvero di rinunciare – facendo sacrificio di sé-- al suo progetto; e di tornare. Ma non in Istituto: a casa. Anzi, all'albergo Bologna: la sua casa. Forse pensa davvero di ritornare perché specifica: *ho però intenzione di rinunciare_all'insegnamento.* Non vuole più incontrare la sua bella e vivace allieva Gilda Senatore, di cui diremo?? In ogni caso, anche se tanto ama insegnare, troppo gli peserebbe questo recente, ulteriore compito di comportarsi come gli altri gli chiedono; di porsi sullo stesso piano sul quale tanti altri vivono e *lì* lo vogliono incontrare… <<Troppo era già compreso dall'orrore –scrive Pirandello– di chiudersi nella prigione d'una forma qualunque>>. E si difende: *non mi prendere per una ragazza ibseniana, perché il caso è differente.*

Ma ormai, col prendere la risoluzione che aveva preso la settimana passata, aveva operato il taglio più difficile. E sa guardare alla propria vicenda con sufficiente distacco da dirne: *"il caso* è differente". Si rende conto che, oramai, una concreta speranza l'ha davvero lasciata ai familiari; sì che la madre resterà convinta che il figlio *non* si fosse suicidato; e tale convinzione serbò per tutta la vita, tanto da lasciargli la sua parte nel testamento. E può quindi proseguire col suo programma. <<Non è altro che questo, epigrafe funeraria, un nome. Conviene ai morti. A chi ha concluso – scrive ancora Pirandello--. Io sono vivo e non concludo. La vita non conclude. E non sa di nomi, la vita… Tutto fuori, vagabondo… Così soltanto io posso vivere, ormai. Rinascere per un attimo. Impedire che il pensiero si metta in me di nuovo a lavorare, e dentro mi rifaccia il vuoto delle vane costruzioni…>>.

Senza dimenticare l'ammonizione di E. Fermi: "con la sua intelligenza, una volta che avesse deciso di scomparire, Majorana ci sarebbe riuscito".

### *Dopo:*

Ricordiamo che le testimonianze più serie in nostro possesso paiono suggerire che Ettore se ne andò davvero "in America o altrove": anzi, proprio in Argentina. E nei pressi di Buenos Aires. Esse sono tre, come si è visto, e indipendenti: 1) del Prof. Carlos Rivera, allora direttore dell'Istituto di fisica dell'Università Cattolica di Santiago del Cile; 2) della signora Blanca de Mora, vedova dello scrittore Miguel Asturias, premio Nobel 1967 per la letteratura; 3) e, aggiungiamo qui, del direttore della famosa casa editrice Losada di Buenos Aires. Queste testimonianze, anche se poi non hanno trovato sufficienti riscontri a conferma, ci sono state segnalate non solo dai già nominati illustri fisici Yuval Ne'eman e Tullio Regge, ma anche dalla nota pittrice Carla Tolomeo, e dal critico e scrittore milanese Giancarlo Vigorelli, recentemente mancato. Tutti questi particolari potrebbero essere di poca importanza: degno di maggiore considerazione è il pensiero che, nella sua genialità, Ettore Majorana possa avere scoperto di potere o dover fare qualcosa di più importante, che non il vincere alcuni premi Nobel continuando a restare fra noi.

CAPITOLO 8

**ETTORE MAJORANA: UOMO E SCIENZIATO**

*Il carattere di Ettore*

    Dalle parole di Edoardo Amaldi sappiamo che, dal matrimonio dell'Ing. Fabio con la Sig.ra Dorina Corso, anch'essa di famiglia catanese, erano nati cinque figli: Rosina; Salvatore, dottore in legge e studioso di filosofia; Luciano, ingegnere civile che si dedicò alla progeettzione e costruzione di strumenti per l'astronomia ottica; Ettore (nato il 5 Agosto 1906 a Catania) ; e, quinta e ultima, Maria, musicista .

   Com'era Ettore dal punto di vista umano? Ettore Majorana era persona sensibilissima e introversa, ma profondamente buona. La sua ritrosia e timidezza, e la sua difficoltà di contatto umano —-resa più acuta dalla sua stessa intelligenza—, non gli impedivano di essere affettuoso. Aiutava ad esempio i compagni di studio, al punto di presentarsi a sostenere un esame al posto di un amico timoroso! Era, almeno fino al 1934, di carattere allegro: la sorella Maria, come sappiamo, ne ricordava soprattutto le barzellette, le risate, il gioco alla palla fatto nel corridoio di casa… E tutti i suoi ex-compagni di università ci hanno detto del gran tempo trascorso con Ettore al bar "Il Faraglino" di Roma, o le chiacchierate e discussioni culturali alla "Casina delle Rose" di Villa Borghese.

    Ettore, lo sappiamo, era poi ricchissimo di "humour", cosa abbondantemente confermata da tanti episodi e dal suo epistolario. Consideriamo qui qualcuna delle lettere scritte al compagno ed amico Gastone Piqué: *"corrispondenza", ci disse quest'ultimo, "di carattere prettamente fraterno e di spirito giovanile. Essa rivela il temperamento gioviale e caustico di Ettore, che molti in seguito gli hanno negato".* E aggiunse: *"Ettore era di carattere molto modesto, oltre che naturalmente onesto, altruista e generoso".* Nel 1925, dalla casa di campagna di Passopisciaro, sull'Etna, Ettore scrive al compagno Piqué: <<Caro Gastone,... se non mi viene un accidente, verrò tra pochi giorni. Né devi credere che sia impossibile che mi venga un accidente nel fiore dell'età; al contrario abbilo per molto verosimile. Infatti io sono stato fin dalla nascita un genio ostinatamente immaturo; il tempo e la paglia non sono serviti a nulla e non serviranno mai, e la natura non vorrà essere così maligna da farmi morire immaturamente d'arteriosclerosi...>>. Molto gustosa è la cartolina di un paio d'anni dopo: *Al sig.Gastone Piqué – Via Buonarroti, 42 – Viareggio: "4 agosto 1927 – Caro Gastone: verrò forse domani, nel pomeriggio. Ti*

*cerchèrò attivamente e ordinatamente: 1) al nunero civico 42; 2) al gatto nero; 3) al bagno Felice; 4) sotto quelle finestre. – Saluti affettuosi. – Ettore".* In essa Ettore annuncia all'amico il suo arrivo, indicandogli le località ove sperava incontrarlo. Ci spiegò il destinatario: <<Al numero 42 della via ove abitavo; al "Gatto Nero", caffé della pineta di ponente che frequentavo d'abitudine; al bagno "Felice", ove la mia famiglia ed io avevamo in affitto una cabina; "sotto quelle finestre", alludendo con ciò all'abitazione di una signorina che io allora corteggiavo>>. Piqué non lo disse, ma noi sommessamente –per amore della Storia— qui lo aggiungiamo: quelle finestre appartenevano all'abitazione della figliola di Giovacchino Forzano...

Ma saltiamo alle lettere, molto più tarde, dalla Germania. Leggiamo un brano di un'altra sua lettera del 1933 dall'Istituto per la Fisica Teorica di Lipsia, dopo essere rientrato in Germania da una visita in Italia: <<Cara mamma: ...Io mi trovo benissimo... Ho notato in treno la rigidità di un ufficiale della *Reichswehr* solo nello scompartimento con me, a cui non riusciva di deporre un oggetto sulla rete e quasi di fare il minimo movimento senza sbattere insieme con forza i talloni. Tale rigidità era evidentemente determinata dalla mia presenza, e in realtà sembra che la cortesia squisita ma sostenuta verso gli stranieri faccia parte dello spirito soldatesco prussiano, poiché, mentre egli si sarebbe sentito disonorato se non si fosse precipitato per accendermi una sigaretta, d'altra parte il suo contegno mi ha impedito di scambiare con lui una sola parola all'infuori dei cortesissimi saluti di rigore>>.

L'esperienza in Germania modifica le opinioni di Ettore circa il fascismo e l'incipiente nazismo; probabilmente anche per l'effetto che gli fa il riuscire a vivere da solo (e probabilmente non male) nella bene organizzata e accogliente -- "cortesissima e simpatica"-- città di *Leipzig* . Tanto che, da un lato, è stato detto e sostenuto che Ettore nutrirà poi delle simpatie per il nazismo. Dall'altro lato, però, esiste una testimonianza di un altro grande fisico, Rudolf Peierls, il quale in una lettera ai fratelli Dubini dichiara che verso la fine del 1932 (cioè prima di partire per la Germania, per lo meno ) Ettore era "veramente contrario al fascismo " (*und dem Fashismus sehr entgegengesetzt*). Oltre alla presente testimonianza, al riguardo si può ricordare l'assennato, certamente non sospetto --e per noi conclusivo—, commento di Leonardo Sciascia, quale appare nel suo saggio majoraniano.

### Il curriculum redatto da Majorana

<<*Sono nato a Catania il 5 Agosto 1906. Ho seguito gli studi classici conseguendo la licenza liceale nel 1923; ho poi atteso regolarmente agli studi di ingegneria in Roma fino alla soglia dell'ultimo anno. Nel 1928 , desiderando di occuparmi di scienza pura, ho chiesto ed ottenuto il passaggio alla Facoltà di Fisica e nel 1929 mi sono laureato in Fisica Teorica sotto la direzione di S.E. Enrico Fermi svolgendo la tesi "La teoria quantistica dei nuclei radioattivi" e ottenendo i pieni voti e la lode. Negli anni successivi ho frequentato liberamente l'Istituto di Fisica di Roma seguendo il movimento scientifico e attendendo a ricerche teoriche di varia*

*indole. Ininterrottamente mi sono giovato della guida sapiente e animatrice di S. E. il Professor Enrico Fermi >> .*

Così Ettore Majorana redige il proprio curriculum, nel maggio 1932. Con l'usuale modestia verso di sé (allora aveva già completato, o in corso, i suoi lavori più importanti!). *Liberamente* significa quale assistente volontario, senza una lira di stipendio (e così sarà fino al novembre del 1937: in questo, ovvero nella maniera di venire incontro all'amore per la ricerca dei fisici, l'università italiana può vantarsi di non essere certo migliorata)... Ettore non trascura che a Fermi spettava il titolo di Sua Eccellenza quale membro dell'Accademia d'Italia voluta da Mussolini. Si noti però come egli scriva "*seguendo* il movimento scientifico" e non "*seguendone* il movimento scientifico": come per riferirsi al movimento scientifico mondiale, piuttosto che a quello del gruppo di Fermi... Passeremo ora ad esaminare brevemente –ma lasciando da parte la modestia di Ettore-- l'incredibile opera scientifica del Majorana.

CAPITOLO 9

**L'OPERA SCIENTIFICA**

Abbiamo visto con quale modestia Ettore redige nel maggio del 1932 il proprio curriculum, richiestogli da Fermi onde inoltrare al Consiglio Nazionale delle Ricerche (allora presieduto da Marconi) la domanda per la borsa di studio che deve permettere al Majorana di recarsi a Lipsia da Heisenberg. Modestia resa eclatante dal fatto che allora Ettore aveva già completato, o in corso, molti dei suoi lavori più importanti.

Ettore pubblicò *pochi* articoli scientifici: nove, come le sinfonie di Bethoven (oltre allo scritto semi-divulgativo "Il valore delle leggi statistiche nella fisica e nelle scienze sociali", pubblicato postumo da G. Gentile jr., e del quale riportiamo in Appendice una versione opportunamente ridotta). Ricordiamo che Majorana passò da ingegneria a fisica nel 1928 (anno in cui pubblicò già un articolo, il primo: scritto insieme con l'amico Gentile), e poi si dedicò alla pubblicistica in fisica teorica solo per pochissimi anni, in pratica fino al 1933.

Ma Ettore ci ha lasciato anche numerosissimi manoscritti scientifici inediti. Di essi –dopo averne redatto un catalogo completo— abbiamo finora pubblicato: a) gli appunti delle 16 lezioni universitarie; b) i suoi "Volumetti", appunti "di studio" e ricerca (apparsi, tre anni dopo l'edizione inglese, anche nella lingua originale italiana); c) e soprattutto una selezione dei suoi "Quaderni", appunti di pura ricerca in fisica teorica, apparsi questi ultimi nel 2009 presso la Springer di Berlino. Poiché la leggenda ci parla di un Majorana che scriveva le formule più importanti di una sua scoperta sul pacchetto delle sigarette Macedonia cui era affezionato, accartocciandolo poi e gettandolo via una folta fumata l'ultima delle dieci sigarette, è bene rivelare come l'analisi dei manoscritti mostri invece quanto Ettore fosse diligente e preciso nel lavoro. Tutte le sue scoperte risultano precedute da una indefessa serie di calcoli: anche per i geni, naturalmente, la scienza non può essere solo un semplice gioco di intuizioni. Le formule scarabocchiate sul pacchetto di sigarette (un *vezzo*...) erano, cioè, il sunto di pagine chiare e complete lasciati al sicuro a casa.

*Le prime pubblicazioni*

Torniamo per ora ai soli articoli pubblicati. I primi, redatti tra il 1928 e il 1931, riguardano problemi di fisica atomica e molecolare: per lo più questioni di spettroscopia atomica o di legame chimico (sempre, s'intende, nell'ambito della meccanica quantistica). Come notò Edoardo Amaldi, un esame di questi lavori lascia colpiti per la loro alta classe: essi rivelano sia una profonda conoscenza dei dati sperimentali anche nei più minuti dettagli, sia una disinvoltura non comune, soprattutto a quell'epoca, nello sfruttare le proprietà di *simmetria* degli "stati quantistici" per semplificare qualitativamente i problemi e per scegliere poi la via più semplice per la risoluzione quantitativa.

Tra questi primi articoli ne scegliamo uno solo: "Atomi orientati in campo magnetico variabile" (1932): esso è l'articolo, famoso tra i fisici atomici, in cui viene introdotto l'effetto Majorana-Brossel. Questo lavoro è rimasto anche un classico della trattazione dei processi di ribaltamento dello *spin* (o "spin-flip"). I suoi risultati --una volta estesi, come suggerito dallo stesso Majorana, da Rabi nel 1937 e quindi, nel 1945, da Bloch e Rabi (i quali, entrambi premi Nobel [Rabi: 1944; Bloch: 1952], contribuirono a diffondere quanto trovato da Ettore tredici anni prima)—hanno costituito la base teorica del metodo impiegato ancor oggi, ad esempio, in tutti gli spettrometri a neutroni polarizzati. In questo articolo viene introdotta anche la cosiddetta "sfera di Majorana" (per rappresentare quantità abbastanza complicate come gli spinori quantistici mediante semplici insiemi di punti su di una superficie sferica), di cui ha parlato entusiasticamente --per esempio— il famoso matematico e divulgatore, di Oxford, Roger Penrose.

## I più importanti articoli di Majorana

Gli ultimi tre articolo di Ettore sono tutti di tale importanza che nessuno di essi può restare senza commento.

L'equazione a infinite componenti --- L'articolo "Teoria relativistica di particelle con momento intrinseco arbitrario" (1932) è il tipico esempio di lavoro che precorre talmente i tempi da venire compreso e valutato a fondo solo molti anni dopo.

E' difficile mettere d'accordo meccanica quantistica e relatività... A quel tempo era opinione comune che si potessero scrivere equazioni quantistiche compatibili con la teoria della relatività solo nel caso di particelle a *spin* zero o un mezzo. Convinto del contrario, Ettore comincia a costruire opportune equazioni per i successivi valori possibili per lo spin (uno, tre mezzi, ecc.); finché scopre che si può scrivere un'unica equazione rappresentante un'intera famiglia infinita di particelle a spin qualsiasi: un'idea coraggiosa, oltre che basata su una intelligenza fisica e matematica eccezionali, anche perché allora le particelle note --che ora sono centinaia-- si contavano sulle dita di una mano. Tralascia tutti i singoli casi studiati --senza pubblicarli— e si dedica solo a queste equazioni "a infinite componenti". Alcuni dei suoi risultati vennero riscoperti vari anni dopo, e solo da pochi tra i massimi fisico-matematici del mondo [questi ultimi tutti premi Nobel, naturalmente]. Ettore stesso --pur tanto schivo— riferisce a suo padre da Lipsia il 18 febbraio 1933: <<Nell'ultimo mio articolo pubblicato è contenuta una importante scoperta matematica, come ho potuto accertarmi mediante un colloquio col professor van der Waerden, olandese che insegna qui, una delle maggiori autorità in teoria dei gruppi>>. La teoria del

Majorana è stata reinventata da grandi matematici sovietici negli anni 1948-58, e finalmente applicata dai fisici in anni ancora più tardi. L'articolo iniziale di Ettore, anzi, rimarrà in ombra per ben 34 anni, cioè fino a quando Amaldi lo traduce e segnala al fisico americano D. Fradkin, il quale a sua volta strabilia i teorici delle alte energie rendendo finalmente di pubblico dominio (nel 1966) quanto compiuto da Majorana tanti anni prima. Dalla data del 1966, la fama di Ettore cominciò a crescere costantemente anche tra i fisici delle particelle fondamentali.

Fermiamoci qui, per ora. Nel prossimo capitolo esamineremo rapidamente altri due fondamentali articoli di Ettore Majorana

CAPITOLO 10

## LE ULTIME PUBBLICAZIONI SCIENTIFICHE

Abbiamo visto nel precedente capitolo l'interesse metodologico e l'importanza dei primi sei articoli del Majorana (in particolare del sesto), e –più ancora— le straordinarie scoperte contenute nella prima delle ultime sue *tre* pubblicazioni: tutte eccezionali, e precorritrici dei tempi, tanto da essere state comprese e valorizzate solo decenni dopo, o addirittura in questi anni. Dobbiamo ora rivisitare brevemente le sue due pubblicazioni finali: non prima di avere ricordato ancora una volta l'ingente mole di scritti scientifici inediti lasciatici dal Majorana, privi purtroppo dei suoi ultimi manoscritti, con ogni probabilità i più rilevanti, che sono andati sfortunatamente perduti.

### Gli ultimi due articoli

Non appena, al sorgere del 1932, giunge a Roma notizia degli esperimenti dei Joliot-Curie [premi Nobel 1935 per la chimica], Ettore comprende che essi avevano scoperto il "protone neutro" senza accorgersene. Prima ancora, quindi, che ci fosse l'annuncio ufficiale della scoperta del *neutrone*, effettuata poco dopo da Chadwick [premio Nobel 1935 per la fisica], Majorana è in grado di spiegare la struttura e la stabilità dei nuclei atomici mediante protoni e neutroni. Ma Ettore non volle pubblicarne nulla, né permise a Fermi di parlarne a Parigi agli inizi di luglio; i suoi colleghi ricordano che già prima di Pasqua era giunto alle conclusioni più importanti della sua teoria: che protoni e neutroni fossero legati da forze quantistiche originate semplicemente dalla loro *indistinguibilità* di fronte alle interazioni nucleari; cioè da "forze di *scambio*" delle rispettive posizioni spaziali. Solo dopo che Heisenberg pubblica un analogo (benché difettoso!) articolo sullo stesso argomento, Fermi riesce a indurre Majorana a recarsi a Lipsia presso il grande collega. E, finalmente, Heisenberg sa convincere Ettore a pubblicare, anche se tanto in ritardo, i propri risultati: "Uber die Kerntheorie", lavoro apparso il 3 marzo 1933 su una importante rivista tedesca (allora la lingua ufficiale della fisica era il tedesco, e non l'inglese come oggidì).

Le forze "di scambio" nucleari sono ora chiamate forze di Heisenberg-Majorana. Ettore ne parla al padre, con l'usuale modestia, nella lettera del 18.2.1933: <<Ho scritto un articolo sulla struttura dei nuclei che a Heisenberg è piaciuto molto benché contenesse alcune correzioni a una sua teoria>>. Sempre su questo lavoro scrive pochi giorni dopo, il 22 febbraio, alla madre: <<Nell'ultimo "colloquio", riunione settimanale a cui partecipano un centinaio tra fisici, matematici, chimici, etc., Heisenberg ha parlato della teoria dei nuclei e mi ha fatto molta réclame a proposito di un lavoro che ho scritto qui. Siamo diventati abbastanza amici…>>. Majorana divenne facilmente amico di Heisenberg anche perché questi, oltre che scienziato, era filosofo; così come altri illustri scienziati che Ettore incontrò con piacere a Lipsia (e a Copenaghen).

Probabilmente la pubblicazione sulla stabilità dei nuclei venne subito riconosciuta dalla comunità scientifica --evento raro, come sappiamo, per gli scritti di Ettore-- anche grazie a questa opportuna "propaganda" fattane da Heisenberg, che proprio pochi mesi dopo riceverà il premio Nobel.

L'avversione a pubblicare le proprie scoperte, quando esse fossero risultate, all'esame del suo senso ipercritico, o di carattere non abbastanza generale o espresse in forma matematica non abbastanza stringente ed elegante, divenne per Ettore, lo si sa, anche motivo di vezzo. Abbiamo già ricordato, con Amaldi, che, durante le sue laconiche conversazioni coi colleghi di via Panisperna, in Roma, Ettore a un certo punto tirava fuori dalla tasca il suo pacchetto di sigarette Macedonia sul quale aveva scritto le formule principali di una nuova teoria; copiava sulla lavagna quel tanto che era necessario per chiarire il problema, per poi accartocciare il pacchetto e buttarlo nel cestino. Majorana, è vero, non ne pubblicava nulla: ma sappiamo che a casa teneva ben scritta, su fogli ordinati, la sua spiegazione teorica del fenomeno discusso…

Estremamente interessanti sono pure altri passi di lettera. Il 14.2.1933, sempre da Lipsia, Majorana racconta alla madre: <<L'ambiente dell'istituto fisico è molto simpatico. Sono in ottimi rapporti con Heisenberg, con Hund e con tutti gli altri. Sto scrivendo alcuni articoli in tedesco. Il primo è già pronto…>>. Il lavoro "già pronto" è naturalmente quello sulle forze nucleari di cui si sta parlando; il quale, però, rimase l'unico in lingua tedesca. Ancora: nella lettera del 18 febbraio dichiara al padre <<pubblicherò in tedesco, estendendolo, anche l'ultimo mio articolo…>>, riferendosi a quello (visto nel nostro precedente capitolo) dell'"equazione a infinite componenti". In realtà Ettore *non pubblicò più nulla*, né in Germania, né al rientro in Italia, a parte l'articolo (del 1937) di cui stiamo per dire. Di notevole rilievo è quindi sapere che Ettore stesse scrivendo altri lavori: in particolare, che stesse allora estendendo il suo importantissimo articolo sulla equazione a infinite componenti: cosa ampiamente confermata dai rendiconti mensili che Ettore era obbligato a inviare al Consiglio Nazionale delle Ricerche. Majorana continuerà poi a dedicarsi alla fisica, ed a produrre risultati (a noi ignoti), anche negli anni successivi: come lui stesso dichiara allo zio Quirino in una sua lettera degli inizi del 1936.

*Il neutrino di Majorana*

Dai manoscritti ritrovati pare, come si è detto, che Majorana formulasse in quegli stessi anni (1932-33) le linee essenziali anche della sua teoria simmetrica per l'elettrone e l'anti-elettrone: che le formulasse, cioè, non appena si diffuse la notizia della scoperta dell'anti-elettrone, o "positrone". Anche se Ettore pubblica tale teoria solo quando si accinge a partecipare al Concorso a cattedra del 1937. Questa pubblicazione viene inizialmente notata quasi esclusivamente per alcuni risultati utili, ma di poco conto rispetto agli altri ivi contenuti (altri risultati, per contro, quali l'introduzione del cosiddetto "oscillatore di Majorana", sono conosciuti da alcuni matematici, ma sono tuttora sconosciuti ai fisici).

La conseguenza più notevole di tale teoria, comunque, è che il neutrino possa coincidere con la propria antiparticella (l'antineutrino). Come per altri scritti di Majorana, anche questo articolo ha cominciato ad avere fortuna solo vent'anni dopo, a partire dal 1957. Dopo di che ha goduto di fama via via crescente tra i fisici delle particelle relativistiche. Ora sono di moda espressioni come "neutrini di Majorana", "massa di Majorana", "spinori di Majorana"; e perfino "majoroni". In questo momento, in tutto il mondo, dagli Usa al Giappone e all'Italia (Laboratori del Gran Sasso) sono in corso colossali esperimenti per scoprire se i neutrini reali sono del tipo Majorana, o del tipo comune ("di Dirac"). E a Catania si spera di realizzare un esperimento ancora più colossale: quello con un chilometro cubo di acqua purissima immerso nel mare...

Terminiamo qui le considerazioni scientifiche, per tornare a trattare questioni culturali in senso lato, connesse col nome di Ettore Majorana. Osserviamo soltanto che perfino i lavori dal Majorana *pubblicati* (per non parlare dei suoi manoscritti inediti) sono per la Fisica una miniera parzialmente inesplorata. Solo di recente, ad esempio, è stato notato come nelle prime pagine dell'ultimo suo articolo si trovi una chiara formulazione del "principio d'azione quantistico", che ha portato agli sviluppi recenti più importanti in quella che è nota come teoria dei campi quanto-relativistici.

# CAPITOLO 11

## MAJORANA, SCIASCIA, E LA RESPONSABILITA' DEGLI INTELLETTUALI

Parlando di Majorana non si può trascurare il fatto che Leonardo Sciascia (che ci ha lasciato venti anni fa) dedicò alla di lui scomparsa un saggio, famoso in Italia e nel mondo, intitolato proprio *La scomparsa di Majorana*. Ricordiamo pure che pochi anni or sono, nel 2005, è caduto il trentennale della pubblicazione di tale saggio (mentre il 2006 ha segnato il centenario della nascita di Ettore Majorana).

Sciascia attribuiva la massima importanza proprio a tale suo libro su Majorana.
Ed è bene tornare a valutare criticamente i temi sostenuti o toccati da Sciascia in questo suo volume e negli scritti giornalistici successivi, anche perché si tratta di temi che riguardano da vicino la sempre attuale questione della responsabilità dell'intellettuale nei confronti dell'uso che può essere fatto delle conquiste della tecnologia.

Si è detto che Sciascia ha dato un'importanza via via crescente al proprio saggio sulla scomparsa di Majorana. In un'intervista nella quale a Leonardo Sciascia veniva richiesto di dire quale fosse tra i suoi libri quello che a lui più piacesse, Leonardo rispose: «Fino a qualche anno fa, avrei detto *Morte di un Inquisitore,* ora invece rispondo *La Scomparsa di Ettore Majorana.*». Sciascia-detective non poteva non essere affascinato da quel giallo, di alto livello culturale, quale è la vicenda relativa alla scomparsa dell'eccelso fisico teorico: ma perché tanto interesse, e duraturo nel tempo, da parte di Sciascia per questo personaggio e per le sue vicende del lontano 1938 ?

In un libro dello stesso Sciascia, *Fatti Diversi di Storia Letteraria e Civile* (Sellerio, Palermo), possiamo rileggere quanto da lui scritto in origine su "La Stampa" di Torino per commentare la tarda pubblicazione, da parte di Emilio Segrè, di una discussa lettera indirizzata a quest'ultimo dal Majorana nel 1933. A proposito del proprio racconto "misto di storia e d'invenzione", Sciascia aveva dichiarato:

*«L'avevo scritto, questo racconto, nella memoria che avevo della scomparsa e su documenti che, per tramite del professor Recami, ero riuscito ad avere, dopo aver casualmente sentito un fisico parlare con soddisfazione [il titolo del pezzo giornalistico, qui, è "Majorana e Segré"], ed entusiasmo persino, delle bombe che avevano distrutto Hiroshima e Nagasaki. Per indignazione, dunque; e tra documenti e immaginazione, i documenti aiutando a rendere probante l'immaginazione, avevo fatto di Majorana il simbolo dell'uomo di scienza che rifiuta di immettersi in quella prospettiva di morte cui altri, con disinvoltura a dir poco, si erano avviati ».*

Questo brano già ci rivela la vera ragione dell'interesse costante di Sciascia per l'argomento ivi toccato: ovvero, per la vexata questio della responsabilità della scienza e degli scienziati. Concediamoci due parole di cronaca. Anzitutto, l'incontro di Sciascia con Emilio Segrè (già membro del famoso gruppo romano dei "ragazzi di via Panisperna, guidato da Enrico Fermi) avvenne in Svizzera, presente Moravia, il quale non si peritò di dare qualche gomitata sotto il tavolo a Leonardo quando Segrè cominciò a vantare il suo ruolo nella costruzione della bomba A (e, come vedremo, a Segré non mancava una sua piccola dose di ragioni).

L'agrigentino Sciascia scelse, come contraltare di Segré, il grande conterraneo (catanese) Ettore Majorana –-paragonato da Enrico Fermi a geni come Galilei e Newton-- quale esempio dello scienziato che, di fronte al pericolo che le proprie scoperte possano venire usate a fin di male dal Potere , rinuncia a renderle note, e si ritira nell'ombra. Tale simbolica contrapposizione fu essenzialmente una finzione letteraria; d'altra parte, come Sciascia medesimo scrisse, e come abbiamo già menzionato, il suo racconto è «un misto di storia e di invenzione»: così che, confondendo volontariamente l'essere col dover essere, Sciascia arrivò ad attribuire ad Ettore anche qualità, interessi e decisioni funzionali alla trasformazione della vicenda di Majorana in emblema del comportamento dello scienziato "buono" di fronte ai problemi posti dal progresso scientifico. Ricordiamo, tra parentesi, che nel gruppo di Fermi ci fu davvero chi, sapendo di Los Alamos e della costruzione della bomba, abbandonò la fisica: il grande sperimentale Franco Rasetti. Tralasciata la fisica, divenne un paleontologo di rinomanza internazionale; e, già avanti negli anni, passò poi alla botanica, divenendo alfine uno dei maggiori esperti mondiali di orchidacee.

Esaminando il saggio di Sciascia, si può verificare più di una volta la capacità di *intus legere*, che accompagna l'arte meditata della parola di cui Leonardo Sciascia imbeve i suoi racconti. Leggendo tra le righe, appunto, e in mezzo alle carte, Leonardo seppe intuire alcuni aspetti che sembrano rispondere a verità, come la scoperta di ulteriori documenti negli anni successivi ha parzialmente confermato. Significativa, ad esempio, è la circostanza che Leonardo sostenne che Werner Heisenberg e gli altri scienziati tedeschi *non vollero* accingersi alla costruzione di una bomba atomica: commentando –come noto-- che gli schiavi (di Hitler) si comportarono da liberi... A questa tesi, che raccoglieva ben pochi sostenitori, Sciascia ci teneva; e la conferma di essa è arrivata, eclatante, agli inizi degli anni '90, dopo la dipartita di Sciascia, quando il British Intelligence Service ha tolto il segreto ai "Farm Hall Transcripts". Spieghiamoci. Tra il giugno e il dicembre del 1945 (un periodo che comprese il bombardamento di Hiroshima del 6 agosto), per 24 settimane, dieci tra i più importanti scienziati tedeschi [tra cui Heisenberg, von Weizsaecker, Otto Hahn, Walther Gerlach, Max von Laue] furono tenuti prigionieri nella Farm Hall, presso Cambridge, UK, e le loro conversazioni furono registrate dal servizio segreto britannico a loro insaputa. La traduzione inglese di tali conversazioni (in particolare delle reazioni dei reclusi quando giunse la notizia di Hiroshima e

Nagasaki) è apparsa in istampa nel 1993 nel volume *"Operation Epsilon: The Farm Hall Transcripts"* (I.O.P. Pub.; Bristol). Dalle suddette trascrizioni risulta evidente che, su dieci, solo uno scienziato tedesco (non Heisenberg!) avrebbe voluto, potendo, contribuire alla costruzione della bomba A tedesca.

Significativo è pure il fatto che Sciascia si convinse presto che la scomparsa di Ettore si riferiva a una fuga e non ad un suicidio: ipotesi che sembra la più probabile alla luce dei documenti, pur non decisivi, da noi successivamente rintracciati.

Ebbe poi l'impressione di un latente antagonismo tra Ettore Majorana ed Enrico Fermi, un antagonismo negato da tutti i colleghi e amici di Ettore, ma che, col senno di poi (Ettore abbandonò non solo la famiglia, ma anche il gruppo guidato da Fermi) potrebbe contenere un briciolo di verità. Tale presa di posizione di Sciascia generò, come molti ricorderanno, una vivace polemica tra Leonardo e i fisici, in particolare Edoardo Amaldi; polemica nella quale il sottoscritto prese le parti più di Amaldi che di Sciascia. La polemica riguardò inizialmente quasi solo la questione della partecipazione del Majorana al famoso concorso universitario per la fisica teorica nel 1937 (partecipazione voluta dal gruppo di Fermi –come anche a noi parrebbe-- o decisa da Ettore in contrasto coi colleghi?): ed essa ci vide nella singolare situazione di amici di entrambi i maggiori contendenti, i quali entrambi si sfogavano anche con l'invio di accuminate epistole –l'un contro l'altro armati— al sottoscritto. La polemica presto divenne aspra, tanto che Sciascia arrivò a scrivere (su "La Stampa" della vigilia di Natale del 1975) che "si vive come cani per colpa della scienza": in ciò associandosi un po' pedissequamente a una tradizione di pensiero tipicamente italiana e non molto nobile, che annovera comunque nomi quali il Vico e Benedetto Croce.

Cosa voleva dire Sciascia? Crediamo che lui sapesse che non esistono la scienza o la poesia, ma solo scienziati e poeti; e che le colpe ricadrebbero semmai su (alcuni) scienziati. Crediamo che sapesse, per di più, che, se un poeta o un filosofo pessimisti offrono a un infelice la goccia che lo decide a commettere suicidio, vere colpe non si possano attribuire a quel filosofo o poeta…

Parlando con Sciascia, si era d'accordo nel dire che la colpa dell'esistenza della sedia elettrica non è affibbiabile ad Alessandro Volta; così come la colpa di una rapina a mano armata non è dell'inventore del coltello. Comunque Sciascia ha voluto rinverdire un ricorrente problema, già riproposto in anni non lontani, e in maniera più *soft,* per esempio da Duerrematt, e a proposito del quale proporremo alcune considerazioni: basate in parte sulla constatazione che il problema della potenzialità distruttiva degli strumenti che l'uomo costruisce è vecchio come il mondo. È nato con Prometeo, quando l'uomo ha incominciato a controllare il fuoco. È un problema che ha sentito Alfred Nobel quando, avendo costruito la dinamite (che allevia la fatica delle braccia dell'uomo, ma può divenire arma bellica), creò il premio Nobel, quasi come atto di espiazione.

Ma rileggiamo prima alcune affermazioni di Sciascia, e di Amaldi. In una lettera del 3 dicembre del '75 Amaldi ci dice: «Credo che avrà anche visto l'intervista di Sciascia sul Giornale di Sicilia del 9 novembre del '75. Qui finalmente viene fuori la vera posizione di Sciascia, ossia quella classica in Italia di Croce, di Gentile: la scienza non fa parte della cultura». E Leonardo pochi giorni dopo, il 9 dicembre '75, ci scrive, riferendosi al suo lungo articolo inviato a La Stampa e apparso, come detto, il 24 dicembre del 1975, commentando: «Naturalmente questa è l'ultima volta che scendo in polemica, ed è il caso di dire scendo perché la polemica di Amaldi è piuttosto bassa». In un'altra lettera, il 27 gennaio '76, Leonardo aggiunge infine una affermazione interessante: «Voglio soltanto fare presente che per me l'espressione "rifiuto della scienza" vale "rifiuto della scienza a un certo punto di fronte a certe ricerche, a certe scoperte". E cioè rifiuto da parte degli scienziati stessi».

Abbandoniamo la cronaca, e torniamo al nostro tema principale: il problema della responsabilità degli uomini di scienza.

Premettiamo che Sciascia è uno dei pochissimi scrittori che abbiano parlato di uno scienziato attribuendogli ricchezza e spessore umani e non povertà spirituale; e questo lascia ben sperare per la soluzione del problema annoso delle due culture.

Il problema delle scoperte e invenzioni umane (fosse solo quella del martello) che ammettono applicazioni positive e negative è, dicevamo, un dilemma antico; che ha sentito anche Pierre Curie (il fisico, consorte di Madame Curie), il quale, nel ricevere il premio Nobel per la mitica scoperta del *radium,* ebbe a dichiarare: «Si può pensare che in mani criminali il *radium* possa divenire molto pericoloso, e ci si può chiedere se l'umanità tragga profitto dalla conoscenza dei segreti della natura. L'esempio della scoperta di Nobel [anche Curie lo cita] è caratteristico: Gli esplosivi permettono all'uomo di compiere opere mirabili. Essi sono però anche un mezzo di distruzione in mano ai grandi criminali che spingono i popoli alla guerra. Ma io -- conclude Curie-- sono tra quelli che credono che l'umanità trarrà più bene che male dalle nuove scoperte».

Suggeriamo un'altra considerazione. La costruzione di strumenti è caratteristica *ineliminabile* dell'uomo. Mentre molti animali nell' evoluzione biologica, avendo bisogno per esempio di mascelle più robuste, sviluppano i muscoli della mandibola, l'uomo non fa altrettanto: ma costruisce a partire dalla pietra un coltello di selce. E, se ha bisogno di un braccio più robusto, si limita ad usare un randello; fabbrica, in altre parole, prolungamenti artificiali dei propri organi. E' inevitabile che l'uomo costruisca randelli, e martelli, anche se questi possono essere usati contro i propri simili.

E' forse un problema solo degli scienziati quello del controllo, e dell'uso a fin di bene, delle scoperte e delle invenzioni umane?

Precisiamo alcuni termini della questione. Lo scienziato vero, anzitutto, è quello che fa ricerca solo per amore della conoscenza: per scoprire qualcosa degli elegantissimi segreti della mirabile natura che ci circonda (opera certo non nostra, e, per chi crede, di un Dio infinitamente intelligente). Questo tipo di ricerca --che sempre meno viene finanziata nell'attuale mondo, sensibile quasi solo al denaro-- non può avere limiti, come non può subirli la ricerca poetica. E' invece il tecnologo che si occupa delle eventuali applicazioni dei risultati della ricerca scientifica (anche se lo stesso individuo, in quanto uomo, a un certo punto può smettere i panni dello scienziato per cambiare mestiere, e assumere quelli del tecnologo). Eventuali "colpe" dovrebbero essere attribuite, semmai, ai tecnologi. Ma il tecnologo stesso può giungere alla costruzione, al massimo, di *un unico* prototipo: una primitiva automobile a vapore, ad esempio. E' poi l'intervento del potere economico e politico a determinare la produzione, o meno, di innumerevoli copie del prototipo, e a migliorarlo. Il potere da controllare, pertanto, è quello economico e politico, che spontaneamente tende a ispirarsi al tornaconto, per conseguire il quale alcuni giungono a scatenare o appoggiare guerre economiche e guerre vere. E' ovvio pertanto che questo controllo non può essere demandato alle povere forze degli scienziati, e neppure a quelle dei soli tecnologi: ma esso è compito e dovere di *tutti* i cittadini.

Potremmo interpretare il messaggio di Sciascia nel senso che *anche* gli scienziati devono porsi i problemi che *tutti noi* dobbiamo porci.

Sciascia ci ha ricordato in tal modo la responsabilità che noi tutti abbiamo di fronte all'uso che si fa, nel bene e nel male, delle conquiste del "progresso". E bisogna stare attenti e avvertiti: perché parecchi eredi dell'antico capo tribù, che ora potremmo riconoscere, ad esempio, in alcuni controllori delle grandi società finanziarie internazionali, e anche di varie multinazionali, hanno approfittato della nostra atavica tendenza a colpevolizzare un capro espiatorio, o magari lo stregone della tribù, e si sono riparati dietro la scusa delle necessità del "progresso" e dietro un paravento di accuse ai "moderni stregoni", che hanno contribuito a identificare con gli scienziati… Qualsiasi danno ambientale o eccessivo sfruttamento della natura cominciò così ad essere attribuito alle inevitabili esigenze del progresso associato alla evoluzione della scienza. La cosa prese piede, contribuendo, temiamo, ad avviare il nostro Paese verso un Secondo Medioevo: ma chiaramente non è veritiera.

Abbiamo visto come in realtà non sia soltanto lo scienziato, o non sia soprattutto lo scienziato, ad avere le responsabilità di cui stiamo parlando. E allora, ancora una volta, come possiamo interpretare il messaggio di letterati come Sciascia, che, nel caso di quest'ultimo, lo toccava al punto da fargli attribuire tanta importanza al proprio libro su Majorana? Poniamoci una domanda, questa volta di tipo scientifico-biologico: come mai l'uomo, fra tutti gli animali, è quello che apparentemente è il più feroce coi propri simili? Perché li attacca e tortura, mentre la maggior parte degli animali non si comporta in tal modo? Una ragione biologica c'è; ed è la seguente. Gli animali che nascono con mezzi di offesa scadenti e deboli (come gli uomini, con i loro poveri denti, e unghie) non ricevono in dono dalla natura l'istinto del "cavalierismo" verso il prossimo; mentre gli animali dotati di mezzi di offesa potenti –come lupi o tigri-- posseggono di necessità l'istinto del rispetto intraspecifico: altrimenti la loro specie si sarebbe già estinta! Gli agnelli, per esempio, non hanno denti poderosi, non hanno artigli, e quindi la natura non li ha istintivamente dotati di rispetto per i propri simili; tanto che, trovandosi due agnelli sul ciglio di un burrone, può ben avvenire che uno spinga l'altro giù dal dirupo. Analogamente per due piccioni: essi pure posseggono deboli mezzi di offesa; quando eseguono le loro battaglie mimiche per conquistare il predominio su un territorio, ad un certo punto uno dei due si riconosce perdente, e se ne vola via: e basta. Ma prendete due piccioni maschi e metteteli in una unica gabbia: il vincitore torturerà a morte il perdente... Quando invece sono due lupi a recitare la mimica della loro battaglia (una mimica dalla quale, tra parentesi, nasce la nostra danza tra uomo e donna) onde conquistare il predominio sul branco, a un certo punto uno dei due lupi riconosce la propria inferiorità: questi allora si arrende, e offre il collo, esponendo la giugulare, al vincitore. Il vincitore, nonostante dimostri una gran voglia di azzannare il soccombente, in realtà è costretto dall'istinto a comportarsi da cavaliere: il primo si arrende e il secondo invariabilmente accetta la sua resa e gli risparmia la vita.

Noi uomini non abbiamo ricevuto *in dono* dalla natura l'istinto di rispettare i nostri simili. Però abbiamo poi costruito coltelli, fucili, le bombe atomiche... Cosa occorre allora? Che il rispetto dei nostri simili ce lo dobbiamo guadagnare con ogni sforzo verso una maturazione morale, che deve crescere man mano che le nostre capacità di offesa artificiali aumentano. Si potrebbe dire che Iddio non ci ha voluto fare un tale tipo di regalo, affinchè ce lo si dovesse guadagnare, liberamemte, con lo sviluppo della nostra coscienza morale (a quanto pare, a Dio interessa più la libertà che non l'obbligo a comportarci bene). Vogliamo pensare che questo sia il grande messaggio di letterati e scrittori come Sciascia. Noi ora abbiamo in mano mezzi potentissimi, come aerei da guerra, bombe micidiali, armi chimiche e batteriologiche; e bombe all'idrogeno: oggi basta schiacciare un bottone per uccidere un milione di uomini. *Quindi* abbiamo l'obbligo di sforzarci –per la sopravvivenza della nostra specie— al fine di conseguire un'alta maturazione morale. Maturazione che deve certamente aver luogo nei fisici, negli scienziati, ora diremmo nei biologi, ma che dobbiamo avere tutti, perché è un compito che l'intera l'umanità deve affrontare: solo l'unione di intenti di tutti i cittadini del mondo può imporre ai veri Potenti di perseguire fini di pace.

Per concludere, se lo spazio ce lo concede, una parola per i giovani. Vorremmo ricordare a chi è ancora studente che la preparazione più importante che dobbiamo chiedere alla scuola per la vita è l'attuazione delle nostre potenzialità ereditarie, lo sviluppo delle nostre facoltà morali, intellettuali e cognitive: più che la preparazione a svolgere uno specifico mestiere. Dopo avere conosciuto alcune fette di mondo, ci sentiamo di asserire che la cultura di base che abbiamo (avevamo?) in Italia, e più in generale in Europa, è probabilmente insostituibile. Quando conoscerete altri Paesi, vi renderete conto quanto la mancanza di una forte cultura di base sia deleteria; è importante «perdere tempo» nello studio: sono proprio le cose che apparentemente non servono a nulla, come la letteratura, la poesia, la filosofia, la scienza intesa come conoscenza del mondo, che formano la mente. C'è chi ora esprime seri timori: ad esempio davanti al pericolo che molte delle nostre nazioni divengano troppo organizzate e dominanti, fino a che prendano il sopravvento i burocrati. Da svariati economisti, di vista miope, che divengono consulenti privilegiati di ministri e capi d'industria in ogni settore, a direttori-manager che hanno il compito di trasformare in aziende financo ospedali, università, enti di ricerca scientifica… Tale eccessiva burocratizzazione la si può combattere solo tenendo presente il bene comune, e vivo l'amore per il pensiero indipendente e per la cultura. Il pensiero, la meditazione, la preoccupazione per i nostri simili e per il mondo ci rendono liberi e maturi; molto più degli studi che ci vogliano insegnare *solo* a divenire ingranaggi utili allo sviluppo economico. Non tutti concorderanno con noi, a questo punto, ma riteniamo, per fare un ulteriore esempio, che occorrerebbe premere affinché venga attuato non solo il diritto allo studio, ma anche il diritto alla ricerca scientifica in senso lato; e, non meno, che bisognerebbe limitare l'istituzione di ricerche scientifiche *finalizzate,* ovvero di finanziamenti privilegiati per investigazioni da cui ci si aspettano risultati concreti a breve; perché le grandi innovazioni non scaturiscono mai da ciò che già che si conosce e si può prevedere. E intendiamo riferirci non soltanto al settore delle scienze esatte (che poi esatte non sono), ma pure a quello delle scienze letterarie e morali.

Terminando, facciamo un passo indietro, e ritorniamo alla circostanza che pure Sciascia ebbe a intuire che Majorana probabilmente si ritirò dalla vita consueta, senza suicidarsi; e in effetti Ettore lasciò la famiglia, e il gruppo dei fisici con cui lavorava, con in tasca il passaporto e almeno una quindicina, o ventina, di migliaia di dollari. Questa ipotesi di fuga e non suicidio ci trova consenzienti, anche in base a documenti successivamente ritrovati: i quali suggeriscono come Ettore si sia probabilmente ritirato in un luogo appartato. A tale proposito appoggiamoci di nuovo –ci sia consentita la ripetizione-- alle parole scritteci dalll'italiana Aurora Bernardini, letterata che opera a San Paolo del Brasile:

«L'ipotesi credibile e fondamentata di una sopravvivenza del Majorana è non solo più generosa, ma più rivoluzionaria o, almeno, più progressista del comodistico suicidio. Scartando a pié pari il luogo comune, secondo il quale il genio dei fisici è precoce e di vita breve, o che un fisico può avere un grande talento nel suo ambito ed essere un imbecille nel resto, stando a quanto si sa di Majorana non rimane che credere che in lui la genialità abbia anticipato la scoperta della sua verità, o della verità tout court che Ivan Iljic di Tolstoi scopre solo prima di morire. Quali sono i momenti veramente vivi della vità? Ognuno ha la sua risposta, quasi sempre in ritardo. Majorana l'avrebbe avuta prima. Sarebbe molto utile, per l'odierna umanità, il suo legato in proposito. Forse ancora più utile che il suo legato in quanto fisico».

Il fascino della vita di Majorana, e il suo insegnamento umano, sono forse  questi: ovvero, quelli di uno scienziato che scopre, ad un certo punto, come più importante del ricevere dieci premi Nobel sia il riuscire semplicemente a *vivere*: come ciascuno di noi.

# APPENDICE A

**GLI APPUNTI DI MAJORANA PER LA PROLUSIONE AL SUO CORSO
DI FISICA TEORICA PRESSO LA REGIA UNIVERSITA' DI NAPOLI**
(LEZIONE INAUGURALE DEL 13 GENNAIO 1938)

*Pubblichiamo in questa prima Appendice gli appunti preparati dal Majorana per la lezione inaugurale del suo Corso di fisica teorica preso l'università di Napoli, ove era stato nominato professore ordinario per meriti eccezionali. Tale Prolusione ebbe luogo il 13 gennaio 1938. Nello scritto che appare in questa pagina traspare l'interesse dello scienziato, non solo per le questioni generali e di fondo che animano la ricerca scientifica, ma anche per il migliore metodo didattico da seguire per trasmettere il sapere agli allievi (per i quali allievi nutriva il più profondo interesse).*

*Una lettura degli appunti di Majorana per la sua Prolusione può riuscire rivelatrice riguardo a vari aspetti del carattere scientifico ed umano del Nostro; avvertiamo solo che in essi ci si riferisce alla fisica classica e alla meccanica quantistica, trascurando in questa prima fase gli aspetti relativistici: aspetti che verranno trattati dal Majorana solo nella seconda parte del corso, come rivelato dagli appunti delle sue ultime sei lezioni recentemente scoperti e pubblicati presso Bibliopolis, Napoli (le precedenti dieci lezioni, quelle allora note, furono pubblicate nel 1987 presso il medesimo editore).*

*I presenti appunti sono stati da noi rinvenuti nel già lontano marzo del 1973, e da noi riportati anche nel nostro volume su "Il Caso Majorana: Epistolario, Testimonianze, Documenti" (Mondadori, Milano, 1987 e 1991; Di Renzo, Roma, 2000 e 2002). [Ci si permetta di ricordare ancora una volta che quasi tutto il materiale riguardante Ettore Majorana (fotografie incluse) è coperto da copyright a favore di Maria Majorana in solido col sottoscritto].*

***Gli appunti di Ettore Majorana per la sua lezione inaugurale*** (Università di Napoli; 13.01.1938, ore 9)

ETTORE MAJORANA:

In questa prima lezione di carattere introduttivo illustreremo brevemente gli scopi della fisica moderna e il significato dei suoi metodi, soprattutto in quanto essi hanno di più inaspettato e originale rispetto alla fisica classica.

La fisica atomica, di cui dovremo principalmente occuparci, nonostante le sue numerose e importanti applicazioni pratiche --e quelle di portata più vasta e forse rivoluzionaria che l'avvenire potrà riservarci--, rimane anzitutto una scienza di enorme interesse *speculativo,* per la profondità della sua indagine che va veramente fino all'ultima radice dei fatti naturali. Mi sia perciò consentito di accennare in prima luogo, senza alcun riferimento a speciali categorie di fatti sperimentali e senza l'aiuto del formalismo matematico, ai caratteri generali della concezione della natura che è accettata nella nuova fisica.

\*\*\*

La **fisica classica** (di Galileo e Newton) all'inizio del nostro secolo era interamente legata, come si sa, a quella concezione *meccanicistica* della natura che dalla fisica è dilagata non solo nelle scienze affini, ma anche nella biologia e perfino nelle scienze sociali, informando di sé in tempi a noi abbastanza vicini tutto il pensiero scientifico e buona parte di quella filosofico; benché, a dire il vero, l'utilità del metodo matematico che ne costituiva la sola valida giustificazione sia rimasta sempre circoscritta esclusivamente alla fisica.

Questa concezione della natura poggiava sostanzialmente su due pilastri: l'esistenza aggettiva e indipendente della materia, e il determinismo fisico. In entrambi i casi si tratta, come vedremo, di nozioni derivate dall'esperienza comune e poi generalizzate e rese universali e infallibili soprattutto per il fascino irresistibile che anche sugli spiriti più profondi hanno in ogni tempo esercitato le leggi esatte della fisica, considerate veramente come il segno di un assoluto e la rivelazione dell'essenza dell'universo: i cui segreti, come gia affermava Galileo, sana scritti in caratteri matematici.

L'*oggettività* della materia è, come dicevo, una nozione dell'esperienza comune, poiché questa insegna che gli oggetti materiali hanno un'esistenza a sé, indipendente dal fatto che essi cadano o meno sotto la nostra osservazione. La fisica matematica classica ha aggiunto a questa constatazione elementare la precisazione a la pretesa che di questo mando aggettivo è possibile una rappresentazione mentale completamente adeguata alla sua realtà, e che questa rappresentazione mentale può consistere nella conoscenza di una serie di grandezze numeriche sufficienti a determinare in ogni punto dello spazio e in ogni istante lo stato dell'universo fisico.

Il *determinismo* è invece solo in parte una nozione dell' esperienza comune. Questa dà infatti al riguardo delle indicazioni contraddittorie. Accanto a fatti che si succedono fatalmente, come la caduta di una pietra abbandonata nel vuoto, ve ne sono altri --e non solo nel mondo biologico-- in cui la successione fatale è per lo meno poco evidente. Il determinismo in quanto principio universale della scienza ha potuto perciò essere formulato solo come generalizzazione delle leggi che reggono la meccanica celeste. E` ben noto che un *sistema* di punti --quali, in rapporto alle loro enormi distanze, si possono considerare i corpi del nostro sistema planetario-- si muove e si modifica obbedendo alle leggi di Newton...

*(omissis)*... Ne segue che la configurazione futura del *sistema* può essere prevista con il calcolo purché se ne conosca lo stato iniziale (cioè l'insieme delle posizioni e velocità dei punti che lo compongono). E tutti sanno con quale estremo rigore le osservazioni astronomiche abbiano confermato l'esattezza della legge di Newton; e come gli astronomi siano effettivamente in grado di prevedere con il suo solo aiuto, e anche a grandi distanze di tempo, il minuto preciso in cui avrà luogo un'eclisse, o una congiunzione di pianeti o altri avvenimenti celesti.

<p style="text-align:center">***</p>

Per esporre la **meccanica quantistica** nel suo stato attuale esistono due metodi pressoché opposti. L'uno è il cosiddetto metodo storico: ed esso spiega in qual modo, per indicazioni precise e quasi immediate dell'esperienza, sia sorta la prima idea del nuovo formalismo; e come questo si sia successivamente sviluppato in una maniera obbligata assai più dalla necessità interna che non dal tenere conto di nuovi decisivi fatti sperimentali. L'altro metodo è quello matematico, secondo il quale il formalismo quantistico viene presentato fin dall'inizio nella sua più generale e perciò più chiara impostazione, e solo successivamente se ne illustrano i criteri applicativi. Ciascuno di questi due metodi, se usato in maniera esclusiva, presenta inconvenienti molto gravi.

E un fatto che, quando sorse la meccanica quantistica, essa incontrò per qualche tempo presso molti fisici sorpresa, scetticismo e perfino incomprensione assoluta, e ciò soprattutto perché la sua consistenza logica, coerenza e sufficienza appariva, più che dubbia, inafferrabile. Ciò venne anche, benché del tutto erroneamente, attribuito a una particolare oscurità di esposizione dei primi creatori della nuova meccanica, ma la verità è che essi erano dei fisici, e non dei matematici, e che per essi l'evidenza e giustificazione della teoria consisteva soprattutto nell'immediata applicabilità ai fatti sperimentali che l'avevano suggerita. La formulazione generale, chiara e rigorosa, è venuta dopo, e in parte per opera di cervelli matematici. Se dunque noi rifacessimo semplicemente l'esposizione della teoria secondo il modo della sua apparizione storica, creeremmo dapprima inutilmente uno stato di disagio o di diffidenza, che ha avuto la sua ragione d'essere ma che oggi non e più giustificato e può essere risparmiato. Non solo, ma i fisici --che sono giunti, non senza qualche pena, alla chiarificazione dei metodi quantistici attraverso le esperienze mentali imposte dal loro sviluppo storico-- hanno quasi sempre sentito a un certo momento il bisogno di una maggiore coordinazione logica, di una più perfetta formulazione dei princìpi, e non hanno sdegnato per questo compito l'aiuto dei matematici.

Il secondo metodo, quello puramente matematico, presenta inconvenienti ancora maggiori.. Esso non lascia in alcun modo intendere la genesi del formalismo e in conseguenza il posto che la meccanica quantistica ha nella storia della scienza. Ma soprattutto esso delude nella maniera più completa il desiderio di intuirne in qualche modo il significato fisico, spesso così facilmente soddisfatto dalle teorie classiche. Le applicazioni, poi, benché innumerevoli, appaiono rare, staccate, perfino modeste di fronte alla sua soverchia e incomprensibile generalità.

Il solo mezzo di rendere meno disagevole il cammino a chi intraprende oggi lo studio della fisica atomica, senza nulla sacrificare della genesi storica delle idee e dello stesso linguaggio che dominano attualmente, è quello di premettere un'esposizione il più possibile ampia e chiara degli strumenti matematici essenziali della meccanica quantistica, in modo che essi siano gia pienamente familiari quando verrà il momento di usarli e non spaventino allora o sorprendano per la loro novità: e si possa così procedere speditamente nella derivazione della teoria dai dati dell'esperienza.

Questi strumenti matematici in gran parte preesistevano al sorgere della nuova meccanica (come opera disinteressata di matematici che non prevedevano un così eccezionale campo di applicazione), ma la meccanica quantistica li ha "sforzati" e ampliati per soddisfare alle necessita pratiche; così essi non verranno da noi esposti con criteri da matematici, ma da fisici. Cioè senza preoccupazioni di un eccessivo rigore formale, che non è sempre facile a raggiungersi e spesso del tutto impossibile.

La nostra sola ambizione sarà di esporre con tutta la chiarezza possibile l'uso effettivo che di tali strumenti fanno i fisici da oltre un decennio, nel quale uso -- che non ha mai condotto a difficoltà o ambiguità-- sta la fonte sostanziale della loro certezza. *(Ettore Majorana)*

# APPENDICE B

*L'articolo presentato in questa seconda Appendice fu scritto da Ettore Majorana, in maniera parzialmente didascalica, per una rivista di sociologia; rinunciando poi a pubblicarlo (ed, anzi, cestinandolo). Esso ha visto la luce postumo, per interessamento di Giovanni Gentile jr., grande amico di Ettore, sulla rivista Scientia, vol.36, fascicolo del Febbraio-Marzo del 1942, pp.58-66. Dopo di allora non è stato praticamente più ripubblicato in lingua italiana. Non si sa quando fu scritto: forse negli anni venti, dato che si fa riferimento alla meccanica quantistica standard, senza accenni alle critiche sorte negli anni trenta. Però il tema centrale di questo scritto era ancora vivo nell'animo del Nostro nel 1934: infatti, il 27.07.34 (su carta listata a lutto, dato che quell'anno era mancato suo padre), il Majorana scriverà a Giovannino Gentile di attendersi che <<presto sarà generalmente compreso che la scienza ha cessato di essere una giustificazione per il volgare materialismo>>. Qui presentiamo una riduzione di tale articolo, a nostra cura (inoltre, dato che il testo apparso su Scientia contiene alcuni evidenti errori --commessi nell'interpretazione della calligrafia di Majorana— la presente versione è stata leggermente "editata").*

ETTORE MAJORANA:

**Il valore delle Leggi Statistiche nelle Fisica e nelle Scienze Sociali**

*Riassunto dell'Autore*: La concezione *deterministica* della natura racchiude in sè una reale causa di debolezza nell'irrimediabile contraddizione che essa incontra con i dati più certi della nostra stessa coscienza. G. Sorel tentò di comporre questo dissidio con la distinzione tra *natura artificiale* e *natura naturale* (quest'ultima acausale), ma negò così l'unità della scienza. D'altra parte l'analogia formale tra le leggi statistiche della Fisica e quelle delle Scienze Sociali accreditò l'opinione che anche i fatti umani sottostassero a un rigido determinismo. E' importante, quindi, che i recenti principii della Meccanica Quantistica abbiano portato a riconoscere (oltre ad una certa assenza di oggettività nella descrizione dei fenomeni) il carattere *statistico* anche delle leggi ultime dei processi elementari. Questa conclusione ha reso sostanziale l'analogia tra fisica e scienze sociali, tra le quali è risultata un'identità di valore e di metodo.

**\*\*\*\*\*\*\*\*\*\***

È noto che le leggi della meccanica, in modo particolare, sono apparse lungamente come il tipo insuperabile delle nostre conoscenze della natura, e si è anzi creduto da molti che a tal tipo, in ultima analisi, si sarebbero dovute ricondurre anche tutte le altre scienze. Valga ciò di giustificazione allo studio che intraprendiamo.

# 1. - LA CONCEZIONE DELLA NATURA SECONDO LA FISICA CLASSICA.

Il credito goduto dalla fisica deriva dalla scoperta delle cosiddette "leggi esatte", consistenti in formule relativamente semplici, che si rivelano di universale validità, sia che vengano applicate a nuovi ordini di fenomeni, sia che il progressivo affinamento dell'arte sperimentale le sottoponga a un controllo sempre più rigoroso. È a tutti noto che, secondo la meccanica classica, il movimento di un corpo materiale è *interamente determinato* dalle condizioni iniziali (posizione e velocità) in cui il corpo si trova, e dalle *forze* che agiscono su di esso... In un caso è stato possibile trovare l'espressione generale di queste forze: nel caso cioè che i corpi interagenti siano isolati e agiscano quindi reciprocamente solo *a distanza...*; una situazione di questo tipo la si incontra in presenza della gravitazione universale di Newton, la quale è tipicamente applicabile allo studio dei movimenti degli astri. Come è noto, tale legge è realmente sufficiente per prevedere in ogni aspetto e con esattezza meravigliosa tutto il complesso svolgimento del nostro sistema planetario. [Una sola minuta eccezione, riguardante lo spostamento secolare che subisce il perielio di Mercurio, costituisce una delle maggiori prove sperimentali della recente teoria della relatività generale].

Il successo sensazionale della meccanica applicata all'astronomia ha incoraggiato la supposizione che anche i fenomeni più complicati dell'esperienza comune debbano infine ricondursi a un meccanismo simile, e solo alquanto più generale, della legge di. gravitazione. Secondo tale modo di vedere, che ha dato luogo alla *concezione meccanicistica* della natura, tutto l'universo materiale si svolge obbedendo a una legge inflessibile, in modo che il suo stato in un certo istante è interamente determinato dallo stato in cui si trovava nell'istante precedente; segno che tutto il futuro può essere previsto con assoluta certezza purché lo stato attuale dell'universo sia interamente noto. Tale concezione pienamente deterministica della natura sembra avere avuto in seguito numerose conferme; gli sviluppi ulteriori della fisica classica sembrano vigorosamente confermare il punto essenziale, cioè la completa *causalità* fisica. Non è contestabile che si debba proprio al determinismo il merito principale di aver reso possibile il grandioso sviluppo moderno della scienza, anche in campi lontanissimi dalla fisica. *Eppure il determinismo, che non lascia alcun posto alla libertà umana e obbliga a considerare come illusori, nel loro apparente finalismo, tutti i fenomeni della vita, racchiude una reale causa di debolezza: la contraddizione immediata e irrimediabile con i dati più certi della nostra coscienza.*

Come il suo effettivo e, secondo ogni verosimiglianza, definitivo superamento sia avvenuto proprio nella fisica in questi ultimi anni, diremo più avanti; sarà anzi nostro scopo l'illustrare il rinnovamento che il concetto tradizionale delle leggi statistiche deve subire in conseguenza del nuovo indirizzo seguito dalla fisica contemporanea. Ma per il momento vogliamo ancora attenerci alla concezione classica della fisica, essendo essa ancora la sola largamente conosciuta oltre la cerchia degli specialisti. Prima di chiudere questa parte introduttiva, crediamo opportuno ricordare che le critiche al determinismo si sono nel tempo via via moltiplicate..., invocando alcune volte un principio metafisico di G.B.Vico, e più spesso il principio pragmatista. Quest'ultimo –il principio di giudicare le dottrine scientifiche in base alla loro concreta utilità-- non giustifica in alcun modo, però, la pretesa di condannare l'ideale dell'unità della scienza, che si è rivelata più volte un efficace stimolo al progresso delle idee.

## 2. - IL SIGNIFICATO CLASSICO DELLE LEGGI STATISTICHE E LE STATISTICHE SOCIALI.

Per bene intendere il significato delle leggi statistiche secondo la Meccanica, riferiamoci alla struttura interna dei corpi gassosi, la quale è particolarmente semplice. Infatti, nei gas in condizioni ordinarie le singole molecole xhe li costituiscono si possono considerare come particolarmente indipendenti, e a distanze reciproche considerevoli rispetto alle loro ridottissime dimensioni… [omissis]… Vi è una intera branca della fisica, la termodinamica, i cui principii si possono ricondurre alle nozioni generali della meccanica statistica. Per quanto abbiamo fatto finora, si può così riassumere il significato delle leggi statistiche secondo la fisica classica: l°) i fenomeni naturali obbediscono ad un determinismo assoluto; 2°) l'osservazione *ordinaria* non permette di riconoscere esattamente lo stato *interno* di un corpo, ma solo il suo stato macroscopico; 3°) stabilite delle ipotesi plausibili… il calcolo delle probabilità permette la previsione più o meno certa dei fenomeni futuri.

Possiamo ormai esaminare il rapporto che passa fra le leggi stabilite dalla meccanica classica e quelle regolarità empiriche che sono note con lo stesso nome in modo particolare nelle scienze sociali. L'analogia formale, infatti, non potrebbe essere più stretta. Quando si enuncia, ad esempio, una legge statistica su una certa popolazione, è chiaro che si rinunzia deliberatamente a indagare sulla biografia degli gli individui che compongono la società in esame; non altrimenti, allorché si definisce lo stato (macroscopico) di un gas semplicemente dalla pressione e dal volume, si rinunzia deliberatamente a investigare posizione e velocità di tutte le singole molecole costituenti.

Ammesse le ragioni che fanno credere all'esistenza di una reale analogia fra le leggi statistiche fisiche e sociali, si potrebbe essere indotti a ritenere che, come le prime presuppongono logicamente un rigido determinismo, così le ultime possano essere ritenute da parte loro la prova che il determinismo governa anche i fatti umani; argomento avvalorato dalla tendenza a vedere nella causalità della fisica classica un modello di valore universale.

Sarebbe qui fuor di luogo riprendere discussioni antiche e mai concluse, ma va accolto con viva attenzione l'annunzio che negli ultimissimi anni la fisica è stata costretta ad abbandonare il suo indirizzo tradizionale rigettando, in maniera verosimilmente definitiva, il determinismo assoluto della meccanica classica.

## 3. - LE NUOVE CONCEZIONI .DELLA FISICA.

È impossibile esporre con qualche compiutezza in poche righe lo schema matematico e il contenuto sperimentale della meccanica quantistica: ci limiteremo pertanto a qualche accenno. Vi sono dei fatti sperimentali noti da gran tempo (fenomeni di interferenza) che depongono irrefutabilmente a favore della teoria *ondulatoria* della luce; altri fatti scoperti da recente (effetto Compton) suggeriscono, *al contrario*, non meno decisivamente l'opposta teoria *corpuscolare*. Tutti i tentativi di comporre la contraddizione nel quadro della fisica classica sono rimasti infruttuosi. Senonché di tali fatti inesplicabili, e di molti altri, si è trovata da pochi anni una spiegazione unica e meravigliosamente semplice: quella contenuta nei principii della *meccanica quantistica*.

Gli aspetti caratteristici della meccanica quantistica, in quanto essa si differenzia dalla meccanica classica sono i seguenti:

a) non esistono in natura leggi che esprimano una successione fatale di fenomeni; anche le leggi ultime che riguardano i fenomeni elementari (sistemi atomici) hanno carattere statistico, permettendo di stabilire soltanto la *probabilità* che una misura eseguita su un sistema fisico dia un certo risultato, e ciò qualunque siano i mezzi di cui disponiamo per determinare con la maggior esattezza possibile lo stato iniziale del sistema. Queste leggi statistiche indicano un reale difetto di determinismo, e non hanno nulla di comune con le leggi statistiche classiche... Un esempio ben noto di questo nuovo tipo di leggi naturali è dato da quelle che regolano i processi radioattivi...;

b) una certa mancanza di *oggettività* nella descrizione dei fenomeni. Qualunque esperienza eseguita in un sistema atomico esercita su di esso una perturbazione finita che non può essere, per ragioni di principio, eliminata o ridotta. Il risultato di qualunque misura sembra perciò riguardare piuttosto lo stato in cui il sistema viene portato nel corso dell'esperimento stesso, che non quello inconoscibile in cui si trovava prima di essere perturbato.

La meccanica quantistica ci ha insegnato, come si diceva, a riconoscere che le trasformazioni radioattive sono guidate da una legge elementare non riducibile ad un più semplice meccanismo causale. L'introduzione nella fisica di tale nuovo tipo di leggi probabilistiche, che si nasconde, in luogo del supposto determinismo, sotto le leggi statistiche ordinarie, ci obbliga a rivedere l'analogia che abbiamo più sopra indicato con le leggi statistiche sociali.

È indiscutibile che il carattere statistico di queste ultime deriva almeno in parte dalla maniera in cui vengono definite le condizioni dei fenomeni: maniera generica, cioè propriamente "statistica". D'altra parte, se ricordiamo quanto si è detto più sopra sulle *tavole di mortalità* degli atomi radioattivi, siamo indotti a chiederci se non esista anche qui un'analogia reale con i fatti sociali, che si descrivono con linguaggio alquanto simile.

Bastano comuni artifici di  laboratorio per preparare una catena comunque complessa e vistosa di fenomeni che sia *comandata* dalla disintegrazione accidentale di un solo atomo radioattivo. *Non vi è nulla dal punto·di vista strettamente scientifico che impedisca di considerare come plausibile che all'origine di avvenimenti umani possa trovarsi un fatto vitale egualmente semplice, invisibile e imprevedibile*. Se è così, *come noi riteniamo*, le leggi statistiche delle scienze sociali vedono accresciuto il loro ufficio, che non è soltanto quello di stabilire empiricamente la risultante di un gran numero di cause sconosciute, ma soprattutto di dare della realtà una testimonianza immediata e concreta. La cui interpretazione richiede un'arte speciale, non ultimo sussidio dell'arte di governo. *(Ettore   Majorana -- riduzione dell'A.)*

# APPENDICE C

**LA VITA DI E. MAJORANA: UNA BIOGRAFIA SCHEMATICA**

*In quest'ultima Appendice, per eventuale comodità del lettore, riassumiamo sinteticamente la biografia di Ettore Majorana:*

Ettore Majorana è nato a Catania il 5 agosto di cent'anni fa, quarto di cinque figli, all'interno di una illustre famiglia originaria di Militello Val di Catania. Il padre Fabio Massimo (Catania, 1875; Roma, 1934), ingegnere e fisico, aveva –-tra l'altro-- fondato a Catania la prima impresa telefonica. E' lui ad iniziare l'educazione di Ettore, che fa le prime classi di scuola in casa. Il nonno, Salvatore, era stato due volte ministro nel governo Depretis; quando Ettore nasce, lo zio Angelo (professore universitario di ruolo fin dall'età di 20 anni) sta per la seconda volta ricoprendo la carica di ministro nel governo Giolitti. Un altro zio paterno, Quirino, è un espertissimo fisico sperimentale, professore prima a Torino poi a Bologna, nonché presidente della Società italiana di fisica.

Majorana frequenta quindi medie e ginnasio, a Roma, in collegio, presso l'Istituto parificato M.Massimo dei padri Gesuiti, passando però per l'ultimo anno al Liceo Statale Tasso, e conseguendovi la maturità nel 1923. Lo stesso anno si iscrive ad Ingegneria presso l'Università di Roma, dove ha come compagni il fratello Luciano, Emilio Segré, Giovanni Gentile jr., Giovanni Enriques, Enrico Volterra, Gastone Piqué. Nel 1926 il lungimirante senatore Orso Mario Corbino, siciliano (dal 1918 direttore dell'Istituto fisico di via Panisperna 89A in Roma, e ministro dell'Economia nazionale nel primo governo Mussolini) crea una cattedra di fisica teorica per il venticinquenne Enrico Fermi. Nel giugno del 1927 Corbino rivolge un appello agli studenti di ingegneria affinché, con la nomina di Fermi a Roma, i più portati passino agli studi di fisica; ed Edoardo Amaldi e Segré accettano, portando nel nuovo istituto la notizia delle doti eccezionali di Ettore, e convincendo quest'ultimo, poi, ad incontrare Fermi.

Majorana si trasferisce quindi a Fisica agli inizi del 1928, non senza avere prima sottoposto ad esame le capacità del nuovo maestro. Fermi, lavorando sul modello poi detto di Thomas-Fermi, aveva incontrato un'equazione differenziale che aveva saputo risolvere solo numericamente. Ettore, durante la notte successiva e sfruttando le sue straordinarie capacità pure nel calcolo analitico e numerico, risolve l'equazione analiticamente seguendo due diversi cammini: a) trasformandola in una equazione di Abel; e b) inventando un metodo matematico del tutto nuovo (come hanno recentemente rivelato al nostro collaboratore Salvatore Esposito i suoi appunti dell'epoca). Il giorno dopo verifica che i risultati ottenuti numericamente da Fermi coincidono con i propri; l'esame ha avuto esito positivo. E nel luglio del 1929 il Majorana si laurea in fisica.

Negli anni successivi Ettore si trova inserito nella moderna scuola italiana di fisica, costituita (oltre che da Fermi e, appunto, da Majorana) da teorici come Gian Carlo Wick, Giulio Racah, Giovanni Gentile Jr., Ugo Fano; e da sperimentali come Franco Rasetti, Emilio Segré, Edoardo Amaldi, Bruno Pontecorvo, nonché da un abilissimo chimico. Gli appunti di studio (i cinque "Volumetti") redatti da Majorana tra il 1927 e 11 1932 sono apparsi per i tipi della Kluwer Academic Press nel novembre 2003 in traduzione inglese, e successivamente (2006, Zanichelli) nella lingua originale italiana: essi già rivelano originalità e spirito pionieristico eccezionali. Degli appunti per le sue ricerche (i 18 "Quaderni", oltre a numerosi fogli sparsi), depositati presso la Domus Galilaeana di Pisa [ora, purtroppo, in via di smantellamento, come tante altre prestigiose realtà italiane], una nostra selezione è stata pubblicata nel 2009 a Berlino dalla Springer: essi sono di alto valore, e non solo dal punto di vista storico, costituendo essi, anzi, una ulteriore miniera di idee fisiche e matematiche per la scienza contemporanea.

Il Majorana (M.) consegue la libera docenza in fisica teorica nel novembre del 1932. Nello stesso anno pubblica uno dei suoi articoli più importanti, proponendo per le particelle elementari della materia una equazione "a infinite componenti": è il tipico esempio di lavoro che precorre talmente i tempi da venire compreso solo dopo alcuni decenni; ma, quando viene scoperto, esso rende il M. immediatamente famoso anche tra i fisici delle particelle. Sempre nel 1932, non appena si rende conto per primo che è stato scoperto il neutrone (scoperta che fruttò un premio Nobel), spiega la stabilità dei nuclei atomici senza volere pubblicare nulla. La spiegazione arriva, pubblicata da Werner Heisenberg, ma in forma non del tutto corretta. Fermi, allora, induce M. a compiere una lunga visita nel 1933 a Lipsia, presso Heisenberg, ottenendogli una borsa di sei mesi del C.N.R.; i primi soldi, tra parentesi, guadagnati dal Nostro. E il famoso collega [che pochi mesi dopo riceverà il premio Nobel] convince Ettore a pubblicare la sua teoria, anche se essa costituiva una correzione della propria. Nascono così in fisica nucleare le "forze di scambio" dette di Heisenberg-Majorana.

Durante il periodo di Lipsia il M. aveva già completato altri importanti articoli, andati purtroppo perduti, ma della esistenza di almeno uno dei quali si ha certezza: una teoria completa per le particelle elementari; sarebbe della massima importanza ritrovare i relativi manoscritti, anche se la speranza di avere successo è molto tenue.

Dopo il suo rientro da Lipsia (fine del 1933), il M. dirada sempre più le proprie visite all'Istituto di fisica, continuando però a studiare e lavorare in casa parecchie ore al giorno; e la notte. In una lettera a Quirino del 16.1.1936 il M. informa lo zio di occuparsi, "da qualche tempo, di elettrodinamica quantistica";

conoscendo la sua modestia nell'esprimersi, ciò significa che durante l'anno 1935 il Majorana si era dedicato *a fondo* a ricerche originali, per lo meno in tale settore. Di nuovo, però, non pubblica più nulla, fino alle soglie del concorso a cattedra di cui si sta per dire.

*********************

Dopo il concorso del 1926, in cui aveva ottenuto la cattedra Fermi, passarono altri dieci anni prima che si aprisse, nel 1937, un nuovo concorso per la fisica teorica. Il M., invitato dagli amici e colleghi a mostrare concretamente di essere ancora attivo, trae da un cassetto uno scritto da lui già preparato entro il 1933 (ovvero non appena si era diffusa la notizia della scoperta dell'anti-elettrone) e, nel 1937, lo pubblica: si tratta dell'articolo, ora molto importante (anche dal punto di vista sperimentale), sul "neutrino di Majorana".

I candidati del concorso a cattedre del 1937 sono numerosi, e molti di essi di elevato valore; soprattutto quattro: Ettore Majorana, Giulio Racah, Gian Carlo Wick e Giovanni Gentile Jr. Secondo le regole, solo tre dei partecipanti potevano essere dichiarati vincitori. La commissione giudicatrice, presieduta da Enrico Fermi, dichiara di esitare ad applicare norme concorsuali ad una personalità scientifica della levatura del M., e propone al Ministro dell'Educazione Nazionale, Giuseppe Bottai, di nominare il Majorana professore di Fisica Teorica per alta e meritata fama in una Università del Regno, al di fuori del Concorso. E così avviene; e l'università di destinazione è quella di Napoli.

Il M. prende servizio il 16 novembre 1937 (ottenendo un posto di lavoro retribuito, per la prima volta nella sua vita), tiene la prolusione al corso il 13.01.38, e continua le sue lezioni fino al successivo 23 marzo. Il 24 marzo appare in Istituto, e consegna senza spiegazioni una carpetta di carte (tra le quali non solo i suoi appunti per le lezioni, ma anche articoli scientifici, alcuni già in forma definitiva) alla sua brava e bella allieva Gilda Senatore, successivamente professoressa di fisica. Il giorno successivo scompare. Ciò che era rimasto dei suoi appunti autografi per dieci delle sue lezioni universitarie è stato pubblicato a Napoli nel 1987 presso Bibliopolis: il resto è andato perduto (anche se recentissimamente sono state rinvenuti in Napoli gli appunti delle restanti sei lezioni). Ricordiamo che al nome di Ettore Majorana fin dal 1963 è stato intitolato il noto Centro di Cultura Scientifica, fondato ad Erice (TP) da Antonino Zichichi.

L'epistolario del M., riportato nel nostro libro su Il Caso Majorana insieme con ogni altro documento noto [tutto protetto da copyright] è stato riscoperto da chi scrive nel 1972. Il venerdì 25.03.38, dopo aver lasciato nella camera dell'albergo di Napoli presso il quale aveva preso temporanea residenza una lettera indirizzata "Alla mia famiglia", scrive una prima lettera al direttore, Antonio Carrelli dell'Istituto: in essa, lo informa di avere prese una decisione "oramai inevitabile", nella quale non vi è "un solo granello di egoismo"... Quindi, intascato --sembra-- il passaporto e ritirato lo stipendio relativo ai suoi primi tre (o quattro ) mesi e mezzo di cattedra universitaria, sale sulla nave che faceva servizio tra Napoli e Palermo. Il piroscafo salpa alla 22 e 30'. Tutto fa pensare che egli intenda mettere fine ai propri giorni, o comunque sparire. Invece, il giorno seguente, sabato, sbarca apparentemente a Palermo, nel senso che da lì vengono spediti subito a Carrelli un telegramma urgente, col quale il M. annulla la lettera da Napoli, e una seconda lettera su carta intestata del grand hotel "Sole" di Palermo (la quale costituisce l'ultimo documento autografo rimastoci), nella quale dichiara "il mare mi ha rifiutato", e aggiunge che sta per ritornare a Napoli, anche se con l'intenzione di rinunciare all'insegnamento (non voleva incontrare, timido com'era, la bella e attiva studentessa Gilda?); chiede infine di non volere essere preso "per una ragazza ibseniana perché il caso è differente"...

Il piroscafo riparte la sera da Palermo per Napoli, con arrivo previsto alle ore 5 e 45' del mattino; e Majorana ne acquista un posto di cabina. Tutto lascia ora credere che il M. intenda rientrare a Napoli. Invece, o durante il viaggio, o subito dopo (o subito prima), egli definitivamente scomparve.

Il M. è stato probabilmente il maggior fisico teorico italiano (e forse non solo italiano) del XX secolo. Infatti, quando si ebbe l'impressione che il M. non sarebbe stato più ritrovato, Enrico Fermi, parlando col memore collega Giuseppe Cocconi, non si peritò di affermare che, tra gli scienziati di valore spiccano come vette isolate "i geni, come Galileo e Newton. Ebbene, Ettore era uno di quelli..." E Fermi si espresse in maniera insolita (lui che conosceva praticamente *tutti* i grandi fisici del suo tempo) anche in un'altra occasione, il 27 luglio 1938, scrivendo da Roma al premier del tempo onde chiedere una intensificazione delle ricerche del M.: Fermi sottolinea a Benito Mussolini che "fra tutti gli studiosi italiani e stranieri che ho avuto occasione di avvicinare, il Majorana è fra tutti quello che per profondità di ingegno mi ha maggiormente colpito".

============== FINE ==============